\pdfoutput=1
\documentclass[12pt]{article}
\usepackage[latin9]{inputenc}
\usepackage{geometry}
\usepackage{amsmath}
\usepackage{amssymb}
\usepackage{graphicx}
\usepackage[authoryear]{natbib}
\usepackage{rotating}
\usepackage{booktabs}

\makeatletter
\usepackage{amsfonts}\usepackage{setspace}\usepackage[bottom]{footmisc}\usepackage{indentfirst}\usepackage{endnotes}\usepackage{rotating}\usepackage{caption}\usepackage{tabularx}\usepackage{pdflscape}\usepackage{authblk}
\usepackage[normalem]{ulem}

\usepackage{hyperref}

\def\@biblabel#1{\hspace*{-\labelsep}}

\@ifundefined{showcaptionsetup}{}{%
 \PassOptionsToPackage{caption=false}{subfig}}
\usepackage{subfig}
\makeatother

\begin{document}

{
\singlespacing
\title{\textbf{Gender-Specific Effects of Prenatal Famine Exposure on Educational Attainment: Accounting for Selective Mortality}
\author{
Hiroyuki Kasahara\\
Vancouver School of Economics\\
University of British Columbia\\
hkasahar@mail.ubc.ca 
 \and
 Weina Zhou\\ 
Department of Economics\\
Dalhousie University\\
weina.zhou@dal.ca
}}
\date{February 2026}
\maketitle
}
\begin{center}

 \end{center}
\vspace{-0.2in}
\begin{abstract}

Selective mortality and fertility issues are persistent challenges in estimating the fetal origin effect, with attempts to address these issues being notably scarce. Evidence further suggests that selective mortality is more pronounced in males than in females. This study investigates the causal effects of prenatal exposure to the Great Chinese Famine on educational attainment by addressing gender-specific selection bias. We compare exposed individuals with their unexposed, same-gender siblings, using a famine intensity measure based on county-year level excess death rates. Our findings reveal remarkably similar consequences for both genders: on average, famine exposure increased illiteracy rates by 4 percentage points and decreased years of schooling by 0.3 years for both males and females. These results contribute to our understanding of the long-term impacts of prenatal malnutrition, while accounting for gender-specific selection biases.
\end{abstract}
 
\vspace{0.4cm}


\maketitle

 \newpage

\noindent \doublespacing

\section{Introduction}

The fetal origins hypothesis suggests that conditions experienced in utero, such as maternal nutrition and stress, can have enduring effects on health and economic outcomes in later life.\footnote{See reviews by \cite{2011_Currie_and_Almond}, \cite{2011_Almond}, \cite{2013_Currie_and_Vogl}, \cite{2018_Almond}.  } Despite increased research efforts over the past decade, significant challenges persist in accurately estimating the long-term causal effects of prenatal shocks. A key issue is selective mortality and fertility: individuals with less favorable genetic traits or lower socioeconomic status (SES) may be less likely to be born or survive early childhood. This selection bias complicates efforts to estimate the causal effects of prenatal shocks.

Further complicating these issues, the impact of selection differs between genders. Males in early childhood are more vulnerable to environmental stress than females \citep{Kraemer2000}. It is caused by biological differences, such as differences in hormonal variations and genetic structure (eg. females have 2 X chromosomes, while males have 1 X and 1 Y)  \citep{drevenstedt2008rise}.  The `fragile male' phenomenon is evident in events like the Great Chinese Famine (1959-1961), which caused widespread mortality in rural China \citep{2015_Meng}. Figure 1A illustrates a significant drop in the rural population compared to urban areas during the famine. Figure 1B reveals a pronounced 7 percentage point reduction in the male-female ratio among rural-born individuals during the famine, while the urban ratio remained stable. These data provide concrete evidence that the famine disproportionately affected male births, resulting in a significant number of `missing' males.\footnote{This aligns with \cite{almond2010Chinesefamine}'s observation that individuals born during the Chinese famine are more likely to be female.} 
Notably, this gender-specific selection bias in response to environmental shocks extends beyond severe crises. Even milder shocks, such as increased alcohol availability in Sweden and heightened air pollution in the U.S., have led to reductions in live male births \citep{sanders2015missingmale,nilsson2017alcohol}.

\begin{figure}
\caption{Cohort Size and Gender Ratio from 1953 to 1962 in China }

\center

\includegraphics[scale=0.8]{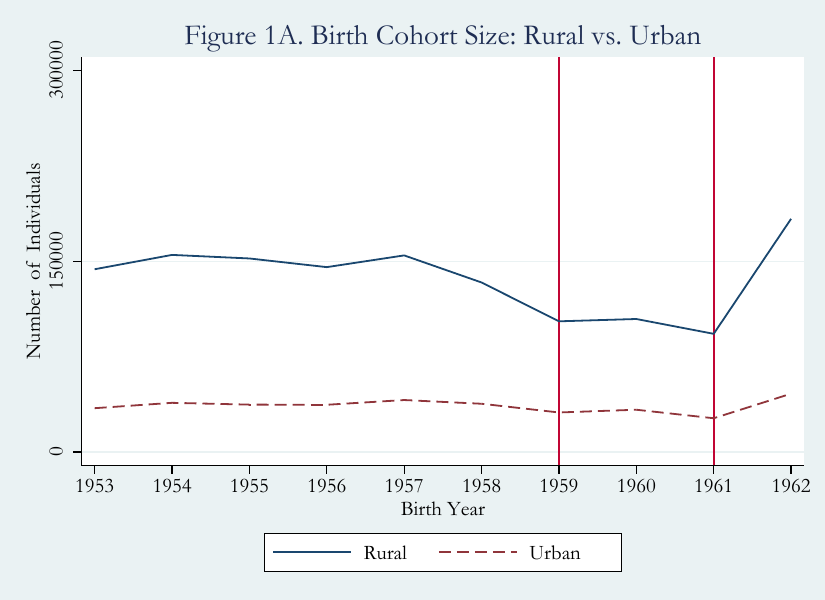}

\subfloat{\includegraphics[scale=0.8]{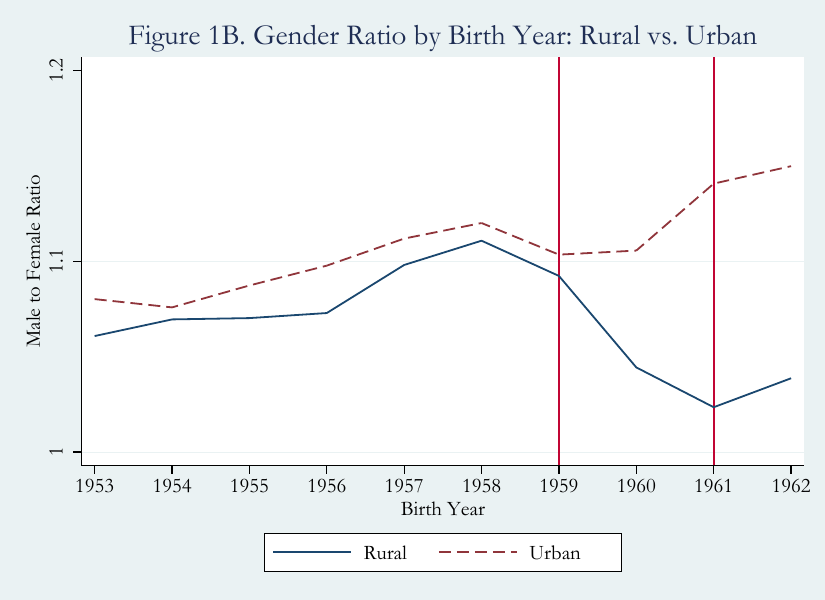}}

\footnotesize
\begin{flushleft}
Notes: Figure 1A presents the cohort size in rural and urban areas,  indicating that famine predominantly occurred in rural areas. Figure 1B compares the gender ratios between famine-affected rural areas and the relatively unaffected urban areas.  An area is defined as rural if 50\% or more of the
population in a county possess rural residential cards (Hukou); others
are defined as urban. Data sources: 1990 population census (1\% sample).
Sample size: 1,788,460.
\end{flushleft}
\end{figure}

The stronger selection bias observed in males compared to females during early childhood may have contributed to the puzzling results in the existing studies. Many studies have found that females exposed to prenatal shocks exhibit worse long-term outcomes compared to similarly exposed males, especially when shocks are severe  \citep{mu2011EHB, chen2013prenatal, akbulut2017war,greve2017fetal}. However, the pronounced gender difference is observed less consistently in contexts involving milder shocks \citep{maccini2009AER, venkataramani2012JHE,van2016instrumental, rosales2019persistent, singhal2019JDE}. This raises a critical question:  Are females inherently more vulnerable to severe prenatal shocks in the long term, or does the stronger selective mortality among males in early life, which results in only healthier males being born and surviving, create the appearance of less severe long-term effects for males? Addressing this question requires a careful examination of selection bias for each gender separately.

This paper aims to fill the gap in the literature by employing a novel \textit{gender-specific}
sibling fixed-effects model to estimate the causal impact of prenatal exposure to the Chinese Famine on educational attainment. 
We deviate from prior research by acknowledging potential heterogeneity in selection bias across genders \citep{almond2006flu1918,bozzoli2009selectionissue,van2022selective}. Ignoring this issue can lead to biased estimates, as disadvantaged fetuses (e.g., those with lower family socioeconomic status or susceptible genotypes)---particularly males---might be more likely to die in early childhood \citep{valente2015civil, Dagnelie2018Congo}.  

To analyze the selection bias problem in estimating the non-random treatment effect parameter,  consider the case where baseline mortality rates are equal across genders in the absence of an prenatal shock. When such a shock occurs, the mortality response differs by gender, resulting in gender-specific thresholds of endowment (such as genetic quality) required for survival, with males having a higher threshold than females, thereby creating gender-specific selection bias. This selection bias can be understood as an omitted variable problem \citep[see][]{1979_Heckman,1997_Angrist}. The primary sources of omitted variables driving the selection are an individual's endowment and unobserved family background characteristics. If these factors determining survival of an prenatal shock are shared between same-gender siblings, then controlling for gender-specific sibling fixed effects (GS-SFE) effectively minimizes the selection bias.

We also analyze the case of heterogeneous treatment effects, where excess death rates' impact on educational outcomes varies across individuals. In this case, we show that our GS-SFE estimator identifies the average treatment effect on the treated (ATT) \textit{among survivors} for a binary treatment variable.

By addressing the issue of gender-specific differences within families, our findings reveal  similar long-term consequences of famine exposure for both males and females. We find that, on average, the famine increased illiteracy rates by 4 percentage points and reduced years of schooling by 0.3 years among those who survived famine, with no statistically significant differences between genders.

The sibling fixed effects model has been employed in existing literature to address unobserved family background characteristics, but without accounting for gender distinctions. Previous research on the Fetal Origins Hypothesis has primarily relied on \textit{gender-neutral} sibling fixed effects approaches to address endogeneity in shock exposure. For instance, \citet{Aizer2016maternal} compared individuals born to the same mother to control for unobserved characteristics that result in differential exposure to maternal stress during pregnancy.
In contrast, our study utilizes \textit{gender-specific} sibling fixed effects to specifically tackle the issue of selective mortality, with careful consideration of gender differences.

Accounting for gender differences is crucial when studying the Chinese Famine. In addition to varying degrees of selective mortality, well-documented son preference in China indicates differential treatment of daughters and sons within families \citep{WeiZhang2011savingrate}. Parental gender stereotypes and cultural norms, which may attribute distinct physical, emotional, or intellectual capacities to sons and daughters, could influence parental behavior differently across genders within a family, invalidating \textit{gender-neutral} sibling fixed effects approaches. By contrast, \textit{gender-specific} sibling fixed effects explicitly address these within-family gender disparities by treating male and female siblings as members of separate family units. 
To further account for possible systematic differences within gender-specific family subunits arising from differential parental treatment among same-gender siblings, we include birth-order fixed effects and control for the number of older brothers and sisters. Additional discussion is provided in Section 5.3, where we assess whether our results could be confounded by intra-household resource reallocation after birth.


China's post-war population expansion policies (1952-1972) provide a unique historical context that enables us to implement a gender-specific sibling fixed-effects approach. These policies, which promoted fertility and restricted abortion, created a generation characterized by large families, offering a natural experiment for within-sibling-gender-unit comparisons. Specifically, according to  the China Statistical Yearbook, the policies doubled China's population from approximately 0.5 billion in 1950 to nearly 1 billion by 1979, when the One-Child Policy was introduced \citep{2014_Zhou}. Consequently, the median family size in our sample of individuals born before the One-Child Policy is four, and the average individual has at least one same-gender sibling.

 

This paper makes a significant contribution to the literature by addressing the persistent empirical challenge of selective mortality and fertility in estimating fetal origin effects. Despite the fundamental importance of addressing selection bias, attempts to do so have been notably scarce in the existing literature. Without adequate controls for gender-specific selection, researchers have encountered paradoxical outcomes. For instance,  \citet*{2012_Goergens}  found that individuals exposed to famine appeared taller than those who were not exposed when selection effects were not controlled for. We demonstrate that a gender-specific sibling fixed-effects model can address such selection bias.

This paper also contributes to existing literature by bridging the
gap between studies that examine the short-term effects of prenatal
shocks on the gender ratio at birth and those that explore the long-term
consequences. A small body of research on the short-term impacts reveals
that prenatal shocks often lead to a decreased proportion of male
births. For instance, alongside evidence from the Chinese Famine \citep{almond2010Chinesefamine},
studies such as \citet{valente2015civil} and \citet*{Dagnelie2018Congo}
demonstrate that civil conflicts increase the likelihood of miscarriages and female live births in Nepal and Congo, respectively. Similar findings have been observed in developed countries. \cite{valente2015civil} further provides a theoretical framework for analyzing gender distortion at birth due to  prenatal  shocks, suggesting that the gender distortion may arise from scarring, selection, or both. \citet{nilsson2017alcohol} utilized Swedish administrative data to show that a temporary increase in alcohol availability among pregnant women is associated with a reduced probability of having a male live birth. Employing a regression discontinuity design, \citet{sanders2015missingmale} found that the Clean Air Act Amendments of 1970 in the US increased the probability of male births, suggesting that prior air pollution reduced male live births. \citet{almond2011ramadon} observed a decrease in the likelihood of live male births among Muslim mothers in Michigan who were pregnant during Ramadan.\footnote{Additionally, \citet{almond2007trivers} reported higher infant mortality rates among male infants born to unmarried and young mothers in the US. The impact of stress on reducing the number of male live births has also been documented following terrorist attacks in Catalonia in 2017 \citep{Bravo2021terrorist} and the 9/11 attacks in New York \citep{Catalano2006_911attack}.}

This paper highlights an important implication drawn from earlier studies: gender-specific \textit{short-term} effects of prenatal shocks at birth likely bias estimates of \textit{long-term} effects differently for each gender. The phenomenon of more males being `missing' due to early-life selection implies that estimation biases in measuring \textit{long-term} effects may be notably more pronounced for males. To our knowledge, this paper is the first to explicitly address gender differences in selection bias when estimating long-term famine effects within a treatment evaluation framework.

This paper has several limitations. First, the long-term effect of famine exposure could be heterogeneous across the endowment distribution. We identify the impact of famine only on survivors, who likely have high endowment quality. The long-term impact of famine exposure on survivors may differ from its potential impact on non-survivors (had they survived) who are likely to have a low endowment. Although the GS-SFE model controls for unobserved endowment quality that could lead to selection bias, we cannot directly test for such \textit{heterogeneity} without data on non-survivors.  However, our estimates from survivors likely provide a useful lower bound for the effects on non-survivors. If famine exposure has a larger impact on disadvantaged individuals---who are less likely to survive---the treatment effects among non-survivors would be larger in magnitude than among survivors, especially for males who have lower survival rates than females.

Second, our estimation strategy relies on the assumption that unobserved endowments can be decomposed into two components: a component shared among same-gender siblings, which is the primary source of selection bias controlled by GS-SFE, and an individual-specific component that varies even among same-gender siblings, which is assumed to be exogenous to the selection process. While the latter exogeneity assumption may be considered strong, it relaxes a typical assumption in the fetal origins literature that within-family variation in endowments is sufficiently small to avoid introducing bias, irrespective of siblings' genders (e.g., Currie and Walker, 2011; Black et al., 2021; Daysal et al., 2022).\footnote{ This assumption is based on the fact that siblings share many family genetic traits and, on average, inherit more than half of the genetic component influencing general cognitive ability (IQ).} Gender differences in mortality responses, even within families, are supported by statistical evidence and biological insights: male siblings inherit an unprotected Y chromosome, increasing their biological vulnerability \citep{Kraemer2000, Ellegren2011sex}; additionally, gender-specific hormones may further contribute to mortality disparities \citep{Bouman2005sex}.\footnote{Gender-specific mortality differences may also be driven by other factors. Some family traits or genetic diseases are passed only to sons, while others are passed only to daughters. These gender-specific traits may result in differences between brothers and sisters in their ability to survive early health shocks and may affect long-term health outcomes differently.} Consequently, endowment differences within same-gender siblings affecting mortality are likely smaller than differences between siblings of different genders, implying that any downward bias from selective mortality in our study is likely smaller compared to biases reported in existing literature, especially for males. Indeed, our empirical results suggest that the gender-specific sibling fixed effects model corrects substantially more bias than gender-neutral sibling fixed effects, particularly for males.

Another potential concern is that Chinese parents make compensatory investments in children who experienced early childhood health shocks \citep{leight2017siblingEDCC}. If such mitigating behaviors is the case in our sample, our estimates likely represent a lower bound.  \citet{yi2015EJ} and \citet{leightLiu2020EDCC} find that such compensatory behavior is more common among highly educated mothers in China. Our results are essentially unchanged when this demographic group is excluded (Table \ref{tab:rb2}, Panel D), suggesting the impact of parental responses in our context is likely to be limited.

Due to population expansion policies and the lack of ultrasound technology to detect pregnancy or fetal sex, families in the 1950s and 1960s generally had limited ability to adopt birth control or engage in sex-selective abortion.\footnote{The ultrasound technology became available in China in 1980s.} We also find suggestive evidence that having a child during the famine did not alter family size or sibling gender composition, further implying that parental behavioral responses regarding gender selection were likely to be limited in our study cohort.

One may further have a concern that within sibling differences could be driven by cohort effects, such as changes in China's socioeconomic and educational environment (e.g., the Cultural Revolution (1966-1976) and population control policies of the 1970s). 
To address this concern, our baseline specification controls for birth-year fixed effects and drops individuals born after the One Child Policy (1979). We further validate our identification strategy using a cohort-by-cohort event-study analysis following Chen and Zhou (2007), confirming that our results are driven by neither pre-existing trends nor post-famine recovery policies (Table \ref{tab:trend}). Our results are robust when adding time-varying regional controls and/or restricting the sample to individuals born before the Cultural Revolution (i.e., before 1966, Table \ref{tab:rb}, Panel D).

\section{Background}

In 1958, the Chinese government initiated the Great Leap Forward campaign,
introducing a series of policies aimed at rapidly transforming China
into an industrialized nation. These policies involved the establishment
of collective farms and communal living, while traditional farming
methods were abandoned. Both labor and capital resources were also
directed from the agricultural sector to the industrial sector. The
central government set unrealistic targets for agricultural production,
and the local government, driven by political considerations, often
exaggerated the reports of actual output. As a result, combined with unfavorable weather conditions, a famine emerged in 1959.

The famine (1959-1961) resulted in an estimated loss of 16.5 to 30
million lives in rural China \citep{2005_Li,2015_Meng}. Although
food shortages were observed in almost every region, food availability
was much worse in rural areas than urban areas, largely due to the
Great Leap Forward policies which sharply increased grain procurement
from rural areas to support the urban population.  
Consequently, the
rural population has substantially declined during famine, while the cohort size is stable in urban areas (Figure 1A). Notably, the
decline in rural areas started even from 1958.\footnote{It is unclear, however, whether the observed decline in the size of
the 1958 birth cohort is due to fewer births or high infant mortality
rates.} Additionally, strict internal migration controls in place before
the famine prevented individuals from escaping regions severely affected
by hunger. \citet{2020_Kasahara} further suggest that an increase
in grain exports during the famine likely worsened the situation.

The average death rate before the famine was 12.1 per thousand during
1954-1958; this number more than doubled in 1960 to 25.54, when the
famine was most severe. As the severity of the famine became increasingly
evident, the central government began to acknowledge the failures
of the Great Leap Forward campaign and reversed many policies introduced
during that period. The recovery from the famine was rather quick:
in 1962, the death rate dropped to 10.31 per thousand, similar to
the pre-famine era.

A growing number of studies have investigated the effect of prenatal
exposure to famine on later-life outcomes. Previous research indicates
that early-life famine exposure is associated with lower height \citep{2007_Chen,2009_Meng_TECH_REPORT,2010_Fung_and_Ha};
anemia \citep{2013_Shi}, the prevalence of hyperglycemia \citep{2010_Li},
high blood pressure \citep{2012_Wang}, overweight or obesity \citep{2006_Luo,2009_Meng_TECH_REPORT},
risk of developing schizophrenia \citep{2005_StClair,2009_Xu}, lower
education and marriage outcome \citep{almond2010Chinesefamine}. While
many studies approximate famine intensity using rough measures of
province-level death rates or county-level cohort sizes from the census
data decades after the famine, two recent studies use hunger recall
to minimize potential measurement error \citep{2020_Cui,2022_Deng}.

There is strong evidence that exposure to famine in utero is associated
with poor later-life outcomes among females, but this association
was less clear among males, as summarized by \citep{2022_Chen}. The
salient gender differences are observed over various later-life outcomes,
for example, education attainment \citep{Shi2011famine, mu2011EHB}
and overweight and obesity \citep{chang2018risks,wang2017association}. However, as discussed earlier, existing findings on gender differences may reflect differences in sample selection bias across genders.

\section{Empirical Framework}

Selective fertility and, to some degree, selective mortality may be attributed to unobserved family-level characteristics such as genetic traits and SES. Healthier families or those with higher SES might have chosen to have more children during the famine, and their offspring, regardless of gender, may have had higher survival rates. These family-level selection issues are likely mitigated by controlling for gender-neutral sibling fixed effects, as siblings share the same family background characteristics.

However, due to biological differences between females and males as discussed early, selective mortality likely exhibits further heterogeneity across genders, even within a family.

The plausible existence of gender-specific unobserved characteristics associated with selective mortality due to  prenatal  shocks, supported by empirical evidence discussed in the introduction, necessitates comparing individuals with their same-gender siblings.

\subsection{Gender-specific selection problem and sibling fixed effects model}
\label{sec:model} 

Consider the following regression model to obtain the causal effect of prenatal famine exposure on later life outcomes:
\begin{equation}
Y_{ig}=\beta_{0}+\beta_{g}EDR_{i}+u_{ig},\label{eq:main}
\end{equation}
where \( Y_{ig} \) represents the educational outcome for individual \( i \) with gender \( g \in\{\text{male},  \text{female}\}\). We use \( EDR_{i} \), excess death rate experienced during the prenatal period, to measure individual \( i \)'s famine exposure.\footnote{We employ a gender-neutral EDR to proxy for the latent environmental shock common to all fetuses. Because males may exhibit higher mortality sensitivity to the same level of deprivation, using gender-specific EDR could mechanically inflate the exposure measure for males, thereby attenuating the estimated coefficient for males (see Appendix \ref{app:D}).} We allow the impact of famine, $\beta_{g}$, to differ by gender. 
The term \( u_{ig} \) captures individual \( i \)'s unobserved family characteristics, such as genetic traits, which could affect \( i \)'s outcomes. We define early childhood as encompassing both the prenatal period and infancy. Unobserved individual characteristics may correlate with both educational outcomes in adulthood and the probability of survival through early childhood. This correlation potentially introduces a sample selection problem when estimating the effect of prenatal shocks on educational outcomes using only the sample of individuals who survived to adulthood.

To explicitly analyze this selection problem, we consider the selection equation during the famine:
\begin{equation} S_{ig} = \mathbf{1} \left\{ \gamma_{0} + \gamma_{g} EDR_{i} + v_{ig} \geq 0 \right\}, \label{eq:selection} \end{equation}
where \( S_{ig} = 1 \) if individual \( i \) of gender \( g \) is observed in the sample, and \( S_{ig} = 0 \) if individual \( i \) is not observed in the sample due to death in early childhood.  
The error term $v_{ig}$, which varies by gender, represents unobserved family or individual characteristics that could affect the individual's survival rate. If individual traits that lead to higher survival rates also tend to have higher educational outcomes, we expect $\text{Cov}[u_{ig}, v_{ig}] > 0$. 

We further allow mortality responses to the prenatal shock to vary by gender, where $\gamma_{g}$ denotes the gender-specific effect on survival rates.\footnote{\citet{valente2015civil} discussed that gender distortions at birth may arise from two factors: \textit{scarring} and \textit{selection}. Scarring occurs when an prenatal shock shifts an individual's endowment distribution to the left, whereas selection occurs when the shock raises the threshold for survival in terms of endowment quality (illustrated as a shift from $d_0$ to $d_1$ in Figure 1 of Valente, 2015). To facilitate a treatment-effect framework, we integrate these conceptual factors into a single selection equation, where $v_{ig}$ captures gender-specific endowment differences and $\gamma_{g}$ represents the gender-specific increase in the survival threshold.} We assume that famine exposure (as measured by $EDR_{i}$) is independent of  $u_{ig}$ and $v_{ig}$. This assumption is further discussed in the subsequent subsection, where we provide a more detailed definition of $EDR_i$.

We anticipate that $\gamma_{g} < 0$, as individuals face a higher likelihood of early childhood mortality during severe famine. Moreover, as observed in the famine and other early studies, the effect of famine on mortality is likely more pronounced in males than in females, implying $\gamma_{\text{male}} < \gamma_{\text{female}} < 0$.

Estimating Equation (\ref{eq:main}) by the OLS, we can
only use the observed sample with $S_{ig}=1$. Note that $S_{ig}=1$
if and only if $v_{ig}\geq-\gamma_{0}-\gamma_{g}EDR_{i}$. To see how the sample selection issue is a form of omitted variable bias issue, taking 
the conditional expectation for both sides of (\ref{eq:main}) given
$S_{ig}=1$ and $EDR_{i}$, we have 
\begin{equation}
E[Y_{ig}|EDR_{i},S_{ig}=1]=\beta_{0}+\beta_{g}EDR_{i}+E[u_{ig}|v_{ig}\geq-\gamma_{0}-\gamma_{g}EDR_{i}]. \label{eq:condieq}
\end{equation} 
 Therefore, because $E[u_{ig}|v_{ig}\geq-\gamma_{0}-\gamma_{g}EDR_{i}]$ is omitted from the specification (\ref{eq:main}), the OLS estimator of \( \beta_{g} \) from the sample of survivors with \( S_{ig} = 1 \) is subject to omitted variable bias, and the bias is larger for males than for females. See Appendix \ref{app:A} for details.

To examine how the \textit{gender-specific} sibling fixed effects (GS-SFE) approach mitigates sample selection bias, let $j(i)$ represent the family to which individual $i$ belongs. We decompose the error terms $u_{ig}$ and $v_{ig}$ as: 
\begin{equation}\label{eq:decomposition}
u_{ig} = \alpha_{j(i)g} + \epsilon_{i}\quad\text{and}\quad v_{ig} = \xi_{j(i)g} + \eta_{i},
\end{equation}
where $\alpha_{j(i)g}$ and $\xi_{j(i)g}$ capture gender-specific unobserved family characteristics (including genetic factors and family wealth).  $\epsilon_{i}$ and $\eta_{i}$ are individual-level components that are not captured by the gender-specific family characteristics. We assume  $\epsilon_{i}$ is mean independent of $\eta_i$ given gender-specific family characteristics, i.e., \( E[\epsilon_{i} | \xi_{j(i)g}, \eta_{i}] = 0 \).

Conditioning on unobserved family characteristics and using \( E[\epsilon_{i} | \xi_{j(i)g}, \eta_{i}] = 0 \), we have
\begin{align}
E[Y_{ig} | EDR_{i}, S_{ig} = 1] &= \beta_{0} + \beta_{g} EDR_{i} + E[(\alpha_{{j(i)g}} + \epsilon_{i}) | \eta_{i} \geq -\gamma_{0} - \gamma_{g} EDR_{i} - \xi_{{j(i)g}}] \nonumber\\
&= \beta_{0} + \beta_{g} EDR_{i} + \theta_{{j(i)g}}, \label{eq:main-wz}
\end{align}
where $ \theta_{{j(i)g}}= E[\alpha_{{j(i)g}} | \eta_{i} \geq -\gamma_{0} - \gamma_{g} EDR_{i} - \xi_{{j(i)g}}]$.\footnote{
Note that \( E[\epsilon_{i} | \eta_{i} \geq -\gamma_{0} - \gamma_{g} EDR_{i} - \xi_{{j(i)g}}] = E[E[\epsilon_{i} |\xi_{j(i)g}, \eta_{i}] | \eta_{i} \geq -\gamma_{0} - \gamma_{g} EDR_{i} - \xi_{{j(i)g}}]=0 \) when $E[\epsilon_{i} |\xi_{j(i)g}, \eta_{i}] =0$.} In this context, we can consistently estimate \( \beta_{g} \) by controlling for \textit{gender-specific} sibling fixed effects \( \theta_{j(i)g} \), using a sample of families with multiple siblings of the same gender \( g \).

Even if mean independence of $\epsilon_i$ from $\eta_i$ conditional on $\xi_{j(i)g}$ does not hold, controlling for GS-SFE likely \textit{reduces} selection bias when gender-specific unobserved family characteristics predominantly drive the bias---that is, when $\text{Cov}[\alpha_{j(i)g} + \epsilon_i,\xi_{j(i)g} + \eta_i]$ is positive and substantially larger than $\text{Cov}[\epsilon_i,,\eta_i]$ (see the discussion in Appendix \ref{app:B}). In this case, our gender-specific fixed effects estimator provides a lower bound on the true effect.

The gender neutral sibling fixed effect model commonly used in the existing literature typically assumes that genetic variation within families is not strong enough to cause systematic differences in outcomes. It implicitly assumes that the primary source of endogeneity arises from family-level unobserved characteristics shared among \textit{all} siblings. However, because mortality selection systematically varies by gender, controlling for  \textit{gender-neutral} sibling fixed effects, rather than  \textit{gender-specific} sibling fixed effects, may be insufficient to address the selection bias. This concern is particularly salient given empirical evidence that famines have differential mortality effects across genders, with males experiencing significantly higher mortality rates than females.
 This paper contributes to the literature by decomposing family-level unobservables into gender-specific family-level components, allowing for a more accurate correction of gender-differentiated selection bias.

The above discussion assumes that the causal parameter $\beta_g$ is non-random and identical across all individuals of the same gender. Suppose instead that the effect of famine on educational outcome $Y_i$ is heterogeneous across individuals. For clarity, we consider a binary treatment indicator $D_i$ 
which equals 1 if $EDR_i$ exceeds its median value and 0 otherwise, and denote the heterogeneous treatment effects by $\beta_{ig}$.  
 In this case, when the mean independence condition $E[\epsilon_i|\eta_i,\xi_{j(i)g}]=0$ holds, comparing individual $i$'s same-gender sibling $i'$ who was born outside the famine period such that $D_{i'}=0$, the gender-specific family fixed effects estimator from regressing $Y_{ig}-Y_{i'g}$ on $D_i$ identifies the average treatment effect on the treated among survivors   (ATT among survivors):
 \begin{equation}\label{eq:ATT-S}
 \hat{\beta}_{g, FE}  \overset{p}{\rightarrow} \frac{E[(D_{i}-E[D_i|S_{ig}=1])D_i  \beta_{ig} | S_{ig}=1] }{ Var(D_i|S_{ig}=1)}=E[\beta_{ig} | D_i=1,S_{ig}=1].
 \end{equation} 
 For further details, see Appendix \ref{app:B}.

Without further assumptions, our setting does not allow us to identify the (unconditional) average treatment effect on the treated (ATT) $E[\beta_{ig}|D_i=1]$ because we cannot identify either the ATT among non-survivors $E[\beta_{ig}|D_i=1, S_{ig}=0]$ or the survival probability $P(S_{ig}=1|D_i)$ from the sample of survivors.

Selection mechanisms may lead individuals with higher survival probability to experience less severe negative treatment effects. For example, individuals with higher genetic endowments are typically more likely to survive adverse conditions $(S_{ig}=1)$, and these same genetic advantages may help them better withstand and recover from the negative effects of famine on later life outcomes. This creates a positive correlation between survival and treatment effects, implying:
$E[\beta_{ig}|D_i=1, S_{ig}=0] < E[\beta_{ig}|D_i=1, S_{ig}=1]$. Under such positive selection, the ATT among survivors $E[\beta_{ig}|D_i=1,S_{ig}=1]$ provides an upper bound for the overall ATT $E[\beta_{ig}|D_i=1]$. Consequently, our negative estimate from the sample of survivors can be interpreted as a lower bound (in absolute value) for the negative effect of famine on educational outcomes among all treated individuals.

\subsection{Empirical Specification}

 Following existing literature \citep{2007_Chen,LI201527}, we use the excess death rate (EDR) to measure the intensity of the famine. $EDR_{yc}$ represents the difference between the death rate in a county $c$ during famine year $y$ and the five-year average death rate preceding the famine period (1954-1958) in the same county.  The death rate is defined as the number of deaths per capita. Following these prior studies, we set the excess death rate to zero during non-famine periods for our baseline estimates.
   
Unlike most early studies that measure EDR at the province-year or prefecture-year levels, our analysis utilizes EDR at the more granular county-year level. Given that a county represents a lower level of aggregation compared to a province or prefecture, our approach allows for a more detailed examination. According to death rate data at the county-year level, sourced from \citet{2020_Kasahara}, over 80\% of the variation in death rates during the famine occurs within provinces. This finer level of measurement of famine intensity facilitates more precise impact estimations in our study.\footnote{China's administrative divisions comprise five levels: province (first
level), prefecture (second level), county (third level), and two smaller
divisions below the county level.}

 To measure an individual's \textit{prenatal} exposure to famine intensity, we weight the excess death rate by the number of months spent in utero during both the birth year and the preceding year. For instance, an individual born in January 1961 is assigned 1/9 of the county's EDR for 1961 and 8/9 for 1960. We name this weighted EDR as Prenatal EDR. This continuous measure captures the variation in exposure intensity more precisely than binary cohort indicators.\footnote{Prenatal EDR varies not only over birth year and county, but also over birth month. For simplicity, we do not explicitly express this variation as an additional subscript in Eq. (\ref{eq:main-1}). The estimation results are essentially unchanged when birth month fixed effects are included in the regression.}
As another example, consider an individual born in February 1959. This individual's nine-month prenatal window spans approximately May 1958 through January 1959. Since we set the excess death rate to zero in non-famine years, only the one month of gestation falling in 1959 contributes to exposure, yielding a Prenatal EDR equal to $1/9$ of the county's EDR for 1959. This example illustrates how the measure naturally handles pregnancies that only partially overlap with the famine period.


We estimate the following regression model: 
\begin{equation}
Y_{ijgyc}=\beta PrenatalEDR_{yc}+\theta_{jg}+\delta_{y}+\epsilon_{ijgyc},\label{eq:main-1}
\end{equation}
where \( Y_{ijgyc} \) represents the education outcome of individual \( i \), with gender \( g \), born to family \( j \), in year \( y \), and in county \( c \). The term \( \theta_{jg} \) denotes gender-specific sibling fixed effects, which effectively treat female and male siblings as belonging to different families. The term \( \delta_y \) captures birth year fixed effects, controlling for factors unique to each year. The error term \( \epsilon_{ijgyc} \) is clustered at the county level, facilitating the handling of within-county correlations.

As discussed in the previous subsection, we compare individuals with their same-gender siblings to address the issue of gender-specific selective mortality. Additionally, the sibling fixed effects help mitigate potential endogeneity related to varying levels of famine exposure. If, within a county, families differ in their ability to access food due to unobserved background characteristics, the sibling fixed effects model effectively controls for these disparities.

One common challenge in estimating a sibling fixed-effects model is the potential for family relocation in response to environmental factors, as noted by \citet{2013_Currie_and_Vogl}. However, in the context of rural China during the 1950s to the 1970s, internal migration was uncommon due to the household registration system \citep{2007_Chen,2017_Wang}.

As discussed in the Introduction section, a concern with our model is that siblings might have differing experiences within the same family-gender unit. To address this, we include birth order fixed effects in our robustness tests, acknowledging that birth order can significantly influence within-family differences \citep{2021_Black}.

Chinese parents may make compensatory investments in children who experienced early childhood health shocks \citep{yi2015EJ, leight2017siblingEDCC}.\footnote{One could potentially test for differential investments using a similar research design, but using data collected when the famine cohorts were young. However, the famine cohort was born half a century ago, and micro survey data in 1960s and 1970s were generally not available.}
Studies also found that parental responses to children's early life conditions are generally more pronounced among highly educated mothers, as opposed to those with lower educational backgrounds \citep{hsin2012Demography, yi2015EJ, leightLiu2020EDCC}. Notably, our study exclusively examines samples from rural areas.  In our robustness checks, we further explore variations across mothers' education levels.

One may have a concern that families could have adopted a stopping rule in their fertility decisions based on the existing number of boys or girls they have. If this stopping rule is affected by the famine, then whether we observe an individual born during the famine could be determined by the gender composition of children born before the famine. Note that the number of existing children born before the famine is fixed at the family level, which is controlled in our estimation. In Section 5.2, we further show that our results are robust against the inclusion of the number of elder brothers or elder sisters, suggesting that this concern is unlikely to bias our results.

Given that the male cohort size is smaller than the female cohort size among those born during the famine, it is possible that some part of gender selection occurs at conception: human beings are more likely to conceive a girl instead of a boy in the presence of distress. If this is the case, girls and boys born into the same family are likely to have similar qualities, and the within-family comparison would be sufficient to address the selective mortality issue. However, it is highly unlikely that the observed gender distortion among the famine birth cohort is mainly driven by gender distortion at conception. \citet{song2012sexratio} documents a statistically significant decline in the sex ratio at birth during the Chinese famine---from 521 to 510 males per thousand births between April 1960 and October 1963---followed by a compensatory rise. This is consistent with the adaptive sex ratio adjustment hypothesis (the Trivers--Willard effect). However, the magnitude of this shift (approximately 2 percentage points) is modest relative to the sharp gender gap in cohort sizes observed during the famine, suggesting that the gender distortion is predominantly driven by differential mortality rather than sex selection at conception.\footnote{\citet{song2010mortality} provides a broader analysis of famine mortality dynamics, including debilitation and selection effects.} While higher male infant mortality and fetal loss in the presence of  prenatal  shocks are widely reported, evidence of large-scale gender distortion at conception remains limited.\footnote{Very few studies are able to examine selection at conception due to the lack of data during early pregnancy, even in developed countries \citep{sanders2015missingmale, valente2015civil}.}

To estimate the gender-specific impacts of prenatal exposure to famine, we estimate:
\begin{equation}
Y_{ijgyc}=\beta_f PrenatalEDR_{yc}\times Female_{i} +\beta_m PrenatalEDR_{yc}\times Male_{i} +\theta_{jg}+\delta_{y}+\epsilon_{ijgyc},\label{eq:main-2}
\end{equation}
where $Female_{i}$ and $Male_{i}$ act as gender indicators, and $\beta_f$ and $\beta_m$ capture the respective famine impacts.

To formally test for gender differences, we employ the following equivalent specification:
\begin{equation}
Y_{ijgyc}=\beta PrenatalEDR_{yc} +\beta_{gd} PrenatalEDR_{yc}\times Male_{i} +\theta_{jg}+\delta_{y}+\epsilon_{ijgyc},\label{eq:main-3}
\end{equation}
where $\beta$ captures the effect on females, and $\beta_{gd}$ explicitly estimates the differential impact on males ($\beta_{gd} = \beta_m - \beta_f$). While mathematically identical to Eq.~(\ref{eq:main-2}), this specification allows us to directly test the statistical significance of the gender gap via the standard error of $\beta_{gd}$.

%
%

\begin{sidewaystable}
    \centering
    \caption{Summary Statistics}\label{table-1}
    \begin{tabular}{l c c c c c c c c c c c}
        \toprule
        & \multicolumn{3}{c}{All} &  & \multicolumn{3}{c}{Male} &  & \multicolumn{3}{c}{Female}  \\ 
        \cmidrule(lr){2-4} \cmidrule(lr){6-8} \cmidrule(lr){10-12}
        &  & Non- &  &  &  & Non- &  &  &  & Non- &  \\      
        & Famine & famine &  &  & Famine & famine &  &   & Famine & famine &  \\
        & cohort & cohort & Difference  &  & cohort & cohort  & Difference  &  & cohort & cohort & Difference \\
        \midrule
        Years of education & 6.70 & 6.36 & 0.34 &  & 7.55 & 6.96 & 0.59 &  & 5.92 & 5.72 & 0.20 \\
        & [4.45] & [4.23] & (0.21) &  & [4.33] & [4.17] & (0.30)** &  & [4.42] & [4.20] & (0.29) \\
        Illiterate & 0.26 & 0.25 & 0.01 &  & 0.20 & 0.21 & 0.01 &  & 0.31 & 0.30 & 0.01 \\
        & [0.44] & [0.43] & (0.02) &  & [0.40] & [0.41] & (0.03) &  & [0.46] & [0.46] & (0.03) \\
        Birth year & 1960.65 & 1965.29 & -4.65 &  & 1960.62 & 1965.02 & -4.40 &  & 1960.67 & 1965.59 & -4.92 \\
        & [1.15] & [7.99] & (0.38)*** &  & [1.17] & [8.08] & (0.55)*** &  & [1.13] & [7.89] & (0.52)*** \\
        Prenatal EDR & 0.05 &  &  &  & 0.05 &  &  &  & 0.05 &  &  \\
        & [0.11] &  &  &  & [0.12] &  &  &  & [0.11] &  &  \\
        \midrule
        Observations & 453 & 4801 & 5254 &  & 218 & 2492 & 2710 &  & 235 & 2309 & 2544 \\
        Number of sibling pairs & 429 & 1645 &  &  & 218 & 1384 &  &  & 231 & 1291 &  \\
        Average number of siblings & 3.14 & 2.72 &  &  & 3.08 & 2.65 &  &  & 3.20 & 2.80 &  \\
        Number of same-gender & \\[-0.8ex]
        \quad sibling pairs & 365 & 1660 &  &  & 177 & 874 &  &  & 188 & 786 &  \\
        Average number of & \\[-0.8ex]
        \quad same-gender siblings & 1.83 & 1.73 &  &  & 1.90 & 1.68 &  &  & 1.84 & 1.80 &  \\
        \bottomrule
    \end{tabular}

    \caption*{Note: The famine cohort is defined as individuals born between January 1959 and August 1962. Prenatal EDR among the non-famine cohort is set to zero. The number of same-gender sibling pairs and the average number of same-gender siblings are computed conditional on individuals having at least one same-gender sibling. Standard deviations are presented in square brackets, while standard errors are reported in parentheses. *** p$<$0.01, ** p$<$0.05, * p$<$0.1.}
\end{sidewaystable}

\section{Data}

Our death rate data are from \citet{2020_Kasahara}, who document
county-level death rates between 1954 and 1965 in 1,803 counties mainly
in rural areas. \citet{2020_Kasahara} collected data from population
statistics yearbooks published by regional statistics bureaus in the
1980s, which were cross-checked with the province-level data used
by \citet{2000_Lin} and \citet{2015_Meng}. The county-year level
EDR data between 1959 and 1961 in \citet{2020_Kasahara}'s whole sample
have a mean of 0.081 with a standard deviation of 0.178. The standard
deviation of EDR after factoring out province fixed effects is 0.15,
suggesting that about 84\% (=0.15/0.178) of the variation in death
rates occurs within provinces.

The underlying mortality data in \citet{2020_Kasahara} are sourced from county gazetteers (\textit{xian zhi}), which are comprehensive local records compiled by county governments. These gazetteers contain annual vital statistics---births, deaths, and population---typically recorded through village-level administrative registries and reported upward through the administrative hierarchy. During the Great Leap Forward and the subsequent Cultural Revolution, administrative disruptions may have affected reporting accuracy, and under-reporting of deaths is a known concern, particularly in the most severely affected areas.

Using the same mortality data form Kasahara and Li (2020), \citet{frank2025sparrows} show that population-weighted means of county-level death rates exhibit near-perfect agreement with province-level rates from Meng et al.\ (2015), with a regression coefficient of approximately 0.996. While this alignment confirms internal consistency rather than external accuracy---since both sources may share reporting biases---the absence of \textit{differential} bias across counties is the relevant concern for our identification strategy, which exploits cross-county variation in famine intensity. The placebo tests reported in Table~\ref{tab:rb} (Panels E--F) further support the interpretation that EDR captures the temporal famine shock rather than persistent institutional deficiencies: falsely assigning famine-era EDRs to unexposed cohorts yields no significant effects.

Our individual-level data is sourced from the China Family Panel Study
(CFPS), a biannual longitudinal survey conducted by Peking University.
This paper utilizes the first wave of the CFPS from 2010 to estimate
the long-term effects of  prenatal  exposure to the famine. A significant
feature of the CFPS 2010 is its comprehensive inclusion of detailed
information about all siblings the respondents have ever had.\footnote{In our sample, 98\% of siblings pairs are born to the same parents, likely due to the extremely low divorce rate in rural China from the 1950s to the 1980s (Yi et al., 2002). Our results remain essentially unchanged when excluding the remaining 2\% of the sample.} Covering 163 counties across 25 provinces,
the CFPS 2010 is considered to be nationally representative.

We matched the data on 1,803 county-year-level famine-era death rates from \citet{2020_Kasahara} with data from 163 counties in the CFPS. Out of these, 52 counties across 18 provinces were successfully matched between the two datasets. Appendix Figure 1 presents the map of provinces with matched counties.\footnote{Since CFPS county identifiers are restricted data, we can only report the provinces with matched counties.} It is important to note that \citet{2020_Kasahara} primarily documents death rates in rural counties, where famines were predominantly experienced (\citealp{2015_Meng}). Consequently, urban areas in the CFPS, such as Shanghai and Guangzhou, which are oversampled, are excluded from our analysis. The average EDR in the matched counties aligns with the overall average of 0.08 reported in \citet{2020_Kasahara}'s full sample. 


Within the 52 matched rural counties, we exclude individuals born before 1949 (the end of the Civil War), those born after the implementation of the One-Child Policy in 1979, and those without siblings. We also exclude individuals whose birth county differs from their current residential county (where the survey was conducted) to address potential measurement error arising from internal migration between 1990 and 2010. Note that ultrasound technology, which facilitated sex-selective abortions, was not available until the early 1980s, after the period covered by our sample.

Table 1 presents the summary statistics. We define the ``Famine cohort" as individuals exposed to the famine (1959--1961) during any month of the nine-month prenatal period. This cohort spans from the oldest individuals born in January 1959 (exposed only during the final month of gestation) to the youngest born in August 1962 (exposed only during the first month). The ``Non-famine cohort'' comprises individuals born before or after this period. The variable ``Illiterate'' is a binary indicator for individuals who did not receive primary education.

Counter-intuitively, the famine cohort exhibits better educational outcomes than the non-famine cohort, particularly among males. Specifically, famine-born males have, on average, 0.59 more years of schooling than their non-famine counterparts, a difference statistically significant at the 5\% level. Famine-born males are also 1 percentage point less likely to be illiterate, though this difference is not statistically significant. These raw patterns stand in contrast to the Fetal Origins Hypothesis, which predicts that prenatal malnutrition leads to worse later-life outcomes. As discussed earlier, this discrepancy likely reflects strong selection bias, where only the healthiest males survived the famine.

%

 The average birth year for the famine cohort is 1960, compared to 1965 for the non-famine cohort. The median family size across the full sample is four. There are 429 sibling pairs that include at least one individual from the famine cohort and 1,645 sibling pairs that include at least one individual from the non-famine cohort. Conditional on individuals having at least one same-gender sibling, the number of same-gender sibling pairs is 365 in the famine cohort and 1,660 in the non-famine cohort. On average, individuals in these same-gender pairs have 1.8 same-gender siblings. The larger family sizes observed in cohorts born before the One-Child Policy are largely attributable to the post-war population boom, driven by policy incentives during the 1950s and 1960s (\citealp{2014_Zhou}).

Finally, we address the trade-off between bias reduction and sample restrictions inherent in our identification strategy, which relies on same-gender sibling pairs. In Appendix \ref{app:C}, we demonstrate that families in our effective sample do not differ significantly in parental education from excluded families (Appendix Table \ref{tab:balance_check}). Furthermore, robustness checks using Inverse Probability Weighting confirm that our results are not driven by specific sample composition (Appendix Table \ref{tab:reweighting}).

\section{Estimation Results}
\subsection{Main Results}
Tables \ref{tab:main-1} and \ref{tab:main-2} present our main findings, with Years of Education and Illiterate serving as the respective outcome variables. Columns 1 and 2 of both tables incorporate gender-specific birth year and county fixed effects (FEs) to account for potential gender differences in these temporal and spatial factors. Column 1 reports the effect of famine exposure on both genders combined. To examine gender-specific effects, we interact $Prenatal EDR$ with gender dummies in column 2. The coefficient on $Prenatal EDR \times Female$ represents the effect on females, while the coefficient on $Prenatal EDR \times Male$ represents the effect on males. 

Without controlling for any sibling fixed effects, the coefficient on $Prenatal EDR$  in column 1 is small and statistically insignificant in both tables. In Table \ref{tab:main-1} column 2, where Years of Education is the dependent variable, the coefficient on the female interaction term is negative and statistically significant, consistent with the Fetal Origin Hypothesis. However, the coefficient on the male interaction term is positive, contrary to the hypothesis. Table \ref{tab:main-2} exhibits a similar pattern, with results suggesting a negative association between the famine exposure and males' illiteracy rate. However, large standard errors render these point estimates statistically insignificant. These results largely align with our model's prediction that males are more susceptible to selection bias than females.

\begin{sidewaystable}
    \centering
    
    \caption{The Long-Term Effects of Prenatal Exposure to the Chinese Famine: Years of Education}
    \label{tab:main-1} 
\begin{tabular}{lcccccccc}
\toprule
 & (1) & (2) & (3) & (4) & (5) & (6) & (7) & (8) \\
\midrule

Prenatal EDR 
    & -1.417 &  & -4.251*** &  & -5.455*** &  & -6.268*** & -5.795*** \\
    & (1.262) &  & (1.103) &  & (1.740) &  & (1.612) & (1.883) \\

Prenatal EDR$\times$Male 
    &  & 1.459 &  & -2.764 &  & -4.824** & 1.444 & 1.173 \\
    &  & (2.380) &  & (1.792) &  & (2.342) & (2.142) & (2.486) \\

Prenatal EDR$\times$Female
    &  & -4.119*** &  & -5.565*** &  & -6.268*** &  &  \\
    &  & (1.073) &  & (1.230) &  & (1.612) &  &  \\

EDR at age 1 
    &  &  &  &  &  &  &  & 1.543 \\
    &  &  &  &  &  &  &  & (2.493) \\

EDR at age 2 
    &  &  &  &  &  &  &  & 3.192 \\
    &  &  &  &  &  &  &  & (2.425) \\

EDR at age 3 
    &  &  &  &  &  &  &  & 1.349 \\
    &  &  &  &  &  &  &  & (1.893) \\

EDR at age 4 
    &  &  &  &  &  &  &  & 1.613 \\
    &  &  &  &  &  &  &  & (2.340) \\

EDR at age 5 
    &  &  &  &  &  &  &  & 2.388 \\
    &  &  &  &  &  &  &  & (1.874) \\

\midrule
Gender-specific Birth Year FE & Yes & Yes & Yes & Yes & Yes & Yes & Yes & Yes \\
Gender-specific County FE     & Yes & Yes &  &  &  &  &  &  \\
Sibling FE (gender neutral)   &  &  & Yes & Yes &  &  &  &  \\
Gender-specific Sibling FE    &  &  &  &  & Yes & Yes & Yes & Yes \\
Observations                  & 5{,}254 & 5{,}254 & 5{,}254 & 5{,}254 & 4{,}155 & 4{,}155 & 4{,}155 & 4{,}055 \\
\bottomrule
\end{tabular}

\caption*{
\small
Notes: Gender-specific birth year, county and sibling FE are fixed effects generated separately for each gender. Sibling FE (gender-neutral) are conventional sibling fixed effects that are the same across genders within a family. Standard errors, clustered at the county level, are reported in parentheses. *** $p<0.01$, ** $p<0.05$, * $p<0.1$.
 }

\end{sidewaystable} 

\begin{sidewaystable}
\centering
\caption{The Long-Term Effects of Prenatal Exposure to the Chinese Famine: Illiterate}
\label{tab:main-2} 
\begin{tabular}{lcccccccc}
\toprule
 & (1) & (2) & (3) & (4) & (5) & (6) & (7) & (8) \\
\midrule

Prenatal EDR 
    & 0.074 &  & 0.469*** &  & 0.694*** &  & 0.829*** & 0.825*** \\
    & (0.130) &  & (0.118) &  & (0.191) &  & (0.283) & (0.251) \\

Prenatal EDR$\times$Male 
    &  & -0.259 &  & 0.403** &  & 0.588** & -0.241 & -0.260 \\
    &  & (0.249) &  & (0.170) &  & (0.229) & (0.326) & (0.304) \\

Prenatal EDR$\times$Female
    &  & 0.386*** &  & 0.527*** &  & 0.829*** &  &  \\
    &  & (0.132) &  & (0.162) &  & (0.283) &  &  \\

EDR at age 1 
    &  &  &  &  &  &  &  & -0.262 \\
    &  &  &  &  &  &  &  & (0.295) \\

EDR at age 2 
    &  &  &  &  &  &  &  & -0.332 \\
    &  &  &  &  &  &  &  & (0.266) \\

EDR at age 3 
    &  &  &  &  &  &  &  & 0.015 \\
    &  &  &  &  &  &  &  & (0.138) \\

EDR at age 4 
    &  &  &  &  &  &  &  & -0.170 \\
    &  &  &  &  &  &  &  & (0.217) \\

EDR at age 5 
    &  &  &  &  &  &  &  & -0.260* \\
    &  &  &  &  &  &  &  & (0.137) \\

\midrule
Gender-specific Birth Year FE & Yes & Yes & Yes & Yes & Yes & Yes & Yes & Yes \\
Gender-specific County FE     & Yes & Yes &  &  &  &  &  &  \\
Sibling FE (gender neutral)   &  &  & Yes & Yes &  &  &  &  \\
Gender-specific Sibling FE    &  &  &  &  & Yes & Yes & Yes & Yes \\
Observations                  & 5{,}254 & 5{,}254 & 5{,}254 & 5{,}254 & 4{,}155 & 4{,}155 & 4{,}155 & 4{,}055 \\
\bottomrule
\end{tabular}

\caption*{
\small
Notes: Gender-specific birth year, county and sibling FE are fixed effects generated separately for each gender. Sibling FE (gender-neutral) are conventional sibling fixed effects that are the same across genders within a family. Standard errors, clustered at the county level, are reported in parentheses. *** $p<0.01$, ** $p<0.05$, * $p<0.1$.
}
\end{sidewaystable}

Columns 3 and 4 refine the analysis from the first two columns by incorporating \textit{gender-neutral} sibling fixed effects. These fixed effects address issues of selective fertility and selective mortality at the family level without distinguishing gender differences within the family. We observe that the estimated effect on females becomes slightly larger in column 4 compared to column 2. Furthermore, the coefficient for males changes direction compared to column 2. The estimated effects for both males and females now align with the prediction of the Fetal Origin Hypothesis. These estimation results indicate that within-family comparisons mitigate selection bias for both genders, at least to some extent. However, the estimated effect on males remains small and is not significantly different from zero in Table \ref{tab:main-1}, possibly because incorporating gender-neutral sibling fixed effects does not fully resolve the selection bias associated with gender-specific selective mortality.

Column 5 and 6 report the results from Equation (\ref{eq:main-1}) and Equation (\ref{eq:main-2}) respectively, where we control for \textit{gender-specific} sibling FEs to account for potential differences in selection bias across genders within a family. Note that gender-specific county FEs and gender-neutral sibling FEs are omitted due to redundancy.

After incorporating gender-specific sibling FEs, the estimated impact of famine exposure on males' Years of Education---captured by the coefficient on $Prenatal EDR \times Male$ (column 6 of Table \ref{tab:main-1})--- becomes considerably larger than in column 4 (-4.824** vs. -2.764), and now is statistically significant at the 5 percent level. 

Column 7 reports the estimated gender difference by including $Prenatal EDR$ and its interaction term with the Male dummy in the regression (Eq\ref{eq:main-3}), where the coefficient on the interaction term formally captures the difference between males and females. The coefficient on interaction term is relatively small (1.444) with a large standard error, suggesting the gender difference is not statistically significant. We observed a similar patten when Illiterate is used as an outcome variable in Table \ref{tab:main-2}. Using the average Prenatal EDR during the famine (0.05), the point estimates in column 7 imply that the famine reduced years of education by approximately 0.3 years (0.05 $\times$ 6.268) and increased the illiteracy by 4 percentage points (0.05$\times$ 0.829), and the gender difference is not statistically significant.

To isolate the impact of famine exposure during the prenatal period from exposure at the other stages of life, column 8 of Table \ref{tab:main-1} and \ref{tab:main-2} includes EDR exposed at ages 1 to 5. We observe that the point estimates of Prenatal EDR and its interaction term essentially unchanged relative to column 7, while the coefficients on EDR at age 1-5 are all statistically insignificant in Table \ref{tab:main-1}. A similar pattern is observed when Illiterate is used as an outcome variable in Table \ref{tab:main-2}, with the only exception being that the coefficient on EDR at age 5 is negative---contrary to our expectation---and statistically significant at the 10\% level. Overall, we conclude that  prenatal  exposure to famine plays a dominate role on educational outcomes compared to the exposure during other early-life stages.

We further report estimation results using a dummy variable to indicate famine intensity in Appendix Table \ref{tab:appendix_binary}, replicating the specifications in column 5-8 of \ref{tab:main-1} and \ref{tab:main-2}. The dummy variable equals one if Prenatal EDR exceeds the median of its positive values, and zero otherwise.  The estimation results using the Prenatal EDR Dummy are all consistent with our main findings.

Existing studies that analyze cross-sectional variations in famine intensity without addressing gender-specific selection bias have found the impact on years of education to be less significant than the effects estimated in our study. For instance, \citet{2009_Meng_TECH_REPORT}  report that prenatal exposure to famine reduces years of education by only 0.012 to 0.016 years for individuals in the 90th percentile of the education distribution, with no average effect observed.\footnote{For individuals in the 90\textsuperscript{th} percentile of the distribution of the outcome, \citet{2009_Meng_TECH_REPORT} report that prenatal exposure to famine reduces those individuals' years of education by 0.174\% in their OLS estimation and 0.238\% in their 2SLS estimation.} \citet{2015_Fan}  observed no significant effect on years of education. Conversely, Almond et al. (2007) found that famine-exposed cohorts were 7.5\% more likely to be illiterate among women, and 9\% more likely among men. Contrary to these findings, our results that account for gender-specific selection bias find substantially larger effects on education, particularly for males.

\begin{table}[h]
    \centering
    \caption{Robustness Checks - Examine Trends Before and After the Famine}
    \label{tab:trend}
\begin{tabular}{l*{4}{c}}
\toprule
 & \multicolumn{2}{c}{Years of Education} & \multicolumn{2}{c}{Illiterate} \\
\cmidrule(lr){2-3}\cmidrule(lr){4-5}
 & (1) & (2) & (3) & (4) \\
\midrule
Prenatal EDR in 1960 $\times$ Born before 1953 & $-1.551$ & $-1.602$ & $-0.160$ & $-0.154$ \\
 & $(1.409)$ & $(1.409)$ & $(0.140)$ & $(0.138)$ \\
Prenatal EDR in 1960 $\times$ Birth 1953--55 & $2.156$ & $2.124$ & $-0.228^{**}$ & $-0.224^{**}$ \\
 & $(1.390)$ & $(1.379)$ & $(0.101)$ & $(0.100)$ \\
Prenatal EDR in 1960 $\times$ Birth 1956--58 & $1.079$ & $0.969$ & $0.063$ & $0.075$ \\
 & $(2.028)$ & $(2.023)$ & $(0.200)$ & $(0.198)$ \\
Prenatal EDR in 1960 $\times$ Birth 1966--68 & $-0.082$ & $-0.140$ & $0.022$ & $0.029$ \\
 & $(0.769)$ & $(0.774)$ & $(0.077)$ & $(0.079)$ \\
Prenatal EDR in 1960 $\times$ Birth 1969--71 & $0.790$ & $0.721$ & $-0.019$ & $-0.011$ \\
 & $(1.056)$ & $(1.053)$ & $(0.100)$ & $(0.102)$ \\
Prenatal EDR in 1960 $\times$ Birth 1972--74 & $-0.664$ & $-0.719$ & $0.075$ & $0.081$ \\
 & $(0.922)$ & $(0.923)$ & $(0.106)$ & $(0.109)$ \\
Prenatal EDR in 1960 $\times$ Born after 1974 & $0.230$ & $0.179$ & $0.029$ & $0.035$ \\
 & $(1.028)$ & $(1.028)$ & $(0.101)$ & $(0.100)$ \\

Prenatal EDR & $-4.704^{***}$ & $-6.074^{***}$ & $0.670^{***}$ & $0.818^{***}$ \\
 & $(1.717)$ & $(1.584)$ & $(0.205)$ & $(0.282)$ \\
Prenatal EDR $\times$ Male &  & $2.467$ & & $-0.267$ \\
 &  & $(2.279)$ & & $(0.315)$ \\
\midrule
Observations & 3,711 & 3,711 & 3,711 & 3,711 \\
\bottomrule
\end{tabular}

\caption*{ 
\footnotesize Notes: The omitted cohort consists of individuals born between September 1962 and December 1965. For the non-famine cohort, prenatal famine intensity is defined as the EDR during the prenatal period that they would have experienced had they been born in 1960. Triple interactions between Prenatal EDR in 1960, the male indicator and non-family cohort dummies are included. Coefficients are reported in  
 Table \ref{tab:triple_interaction}.
Gender-specific birth year and sibling fixed effects are generated separately for each gender.  Standard errors, clustered at the county level, are reported in parentheses. 
${}^{***} p<0.01$, ${}^{**} p<0.05$, ${}^{*} p<0.1$.
 }
  \end{table}

\subsection{Robustness Checks}


We perform several tests to validate our identification strategy and ensure that our results 
are not driven by pre-existing trends or confounding policy shocks. First, following 
\citet{2007_Chen}, we conduct an event-study style analysis to test the parallel trends 
assumption. We interact famine intensity with non-famine birth cohorts to examine whether 
the estimated effects are driven by pre-existing trends or post-famine recovery policies. 
Following the identification strategy in \citet{2007_Chen}, we compare famine-exposed cohorts 
to adjacent non-famine cohorts and test for pre- and post- trends via an event-study design. We further include triple interactions between famine intensity, the male indicator and non-family cohort dummies. This allows us to test whether male-specific trends could affect our results.
 
For individuals in the non-famine cohort, we assign \textit{prenatal} famine intensity as the 
excess death rate (EDR) they \emph{would have experienced had they been in utero in 1960}, 
the peak year of the famine. This counterfactual assignment follows the approach of 
\citet{2007_Chen}.\footnote{Our estimation results are robust to using the average EDR across 
1959--1961 as an alternative measure, rather than relying solely on the 1960 EDR.} 
Because our sample includes individuals born between 1949 and 1979, we group non-famine 
births into three-year cohorts. For the famine cohort, we retain our primary measure of famine 
exposure, \textit{Prenatal EDR}, which weights county-level EDR by the number of months spent 
in utero in each famine year. This measure minimizes measurement error arising from differences 
in gestational timing.


Table \ref{tab:trend} reports the results. The omitted cohort consists of individuals born between September 1962 and 1965. We find that when famine intensity is interacted with non-famine cohorts, the coefficients are generally not statistically significant. The only exception is the cohort born between 1953-1955, where the dependent variable is Illiterate. However, the sign of this coefficient is opposite to that of the prenatal EDR for the famine cohort. Our results are robust when we exclude this group as controls (Appendix Table \ref{tab:appendix_drop1953}). The coefficients on the triple-interaction terms--- Prenatal EDR in 1960 $\times$ Male $\times$ Cohort --- are reported in Appendix Table \ref{tab:triple_interaction}; none of them are statistically significant.

Second, we address concerns regarding potential confounding factors driven by other major historical events or policies. We argue that our Gender-Specific Sibling Fixed Effects (GS-SFE) strategy is particularly robust against such shocks. For a policy to confound our results, it would need to mimic the specific county-by-year intensity of the famine \textit{and} exert a differential impact within same-gender sibling pairs that is proportional to the famine's impact. We believe this specification effectively isolates the famine shock from broader regional policies. Our baseline specification excludes individuals born after the introduction of the One Child Policy (1979) and controls for birth-year fixed effects to account for differences across birth cohorts. In Table 5, Panel A, we incorporate time-varying regional controls---including population size, birth rate, GDP, and province-specific birth-year trends. Panel B further controls for county-year trends. To further address concerns about cohort differences driven by shifts in China's population policy in 1972, the ``Later, Longer, Fewer" policy and change in socioeconomic and educational environment (e.g., the Cultural Revolution (1966--1976)), Panel C drops individuals born after 1972, while Panel D drops individuals born after the Cultural Revolution.

Overall, the estimates in Panels A to D of Table \ref{tab:rb} are consistent with those reported in the last column of Table \ref{tab:main-1} and \ref{tab:main-2}, supporting our main findings.


Third, we conduct placebo tests. Panels E and F falsely assign famine-era EDRs to individuals born a decade before (1949--1951) and after (1969--1971) the famine. This allows us to examine whether the famine treatment could falsely affect siblings who were not exposed to the famine. We do not observe any effects from the falsely assigned EDRs on these cohorts. Crucially, these null results also mitigate concerns regarding systematic measurement error (e.g., political manipulation of mortality records). If high EDRs merely proxied for poor local governance or persistent institutional deficiencies rather than the specific famine shock, we would expect significant negative effects on these unexposed cohorts. The absence of such effects supports the validity of EDR as a measure of the temporal famine shock.

\begin{table}
    \centering
    \caption{Robustness of Identification Strategy}
    \label{tab:rb}
    \small
\begin{tabular}{lccc}
\toprule

\multicolumn{2}{c}{Years of education} & \multicolumn{2}{c}{Illiterate} \\
(1) & (2) & (3) & (4) \\
Prenatal EDR & Prenatal EDR$\times$Male & Prenatal EDR & Prenatal EDR$\times$Male \\
\midrule
\multicolumn{4}{c}{\textbf{\textit{Panel A. Add time variant regional controls}}} \\
$-5.427^{***}$ & 0.904 & $0.700^{***}$ & -0.002 \\
(1.816) & (3.417) & (0.210) & (0.330) \\

\multicolumn{4}{c}{\textbf{\textit{Panel B. Add county-year trend}}} \\
$-5.266^{***}$ & 1.428 & $0.680^{***}$ & -0.020 \\
(1.739) & (3.266) & (0.195) & (0.328) \\

\multicolumn{4}{c}{\textbf{\textit{Panel C. Drop born after 1972}}} \\
$-6.255^{***}$ & 1.641 & $0.876^{***}$ & -0.323 \\
(1.786) & (2.549) & (0.261) & (0.320) \\

\multicolumn{4}{c}{\textbf{\textit{Panel D. Drop born after Cultural Revolution (1966)}}} \\
$-6.979^{***}$ & 2.737 & $1.042^{***}$ & -0.547 \\
(2.006) & (3.075) & (0.263) & (0.375) \\

\multicolumn{4}{c}{\textbf{\textit{Panel E. Placebo test: Assign famine-era EDR}}} \\
\multicolumn{4}{c}{\textbf{\textit{ to individuals born 10 years later}}} \\
1.163 & -0.965 & -0.091 & 0.009 \\
(0.994) & (1.303) & (0.108) & (0.120) \\

\multicolumn{4}{c}{\textbf{\textit{Panel F. Placebo test: Assign famine-era EDR }}} \\
\multicolumn{4}{c}{\textbf{\textit{ to individuals born 10 years earlier}}} \\
-0.717 & 1.130 & -0.192 & 0.295 \\
(1.249) & (2.053) & (0.162) & (0.237) \\

\multicolumn{4}{c}{\textbf{\textit{Panel G. Alternative measure}}} \\
-3.214 & 1.011 & $0.543^{***}$ & -0.334 \\
(2.057) & (3.251) & (0.159) & (0.328) \\

%
%
%
%
%

\bottomrule
\end{tabular}

\caption*{ 
\footnotesize Notes: Gender-specific birth year fixed effects, gender-specific sibling fixed effects, and EDR at ages 1-5 are controlled for in all columns. Each panel reports a single regression, separately estimated using Years of Education and Illiterate as dependent variables, respectively. The reported coefficients reflect the effect of prenatal exposure to famine on both genders (the coefficient on Prenatal EDR), and the differential effect for males relative to females (the coefficient on $Prenatal\ EDR \times Male$), as indicated in the top panel.  The sample sizes are 3,463 (A); 3,463(B); 3,097(C); 1,698(D);  3,964 (E); 4,027 (F); 3,463 (G), respectively.  In Panel A, province-year population size, birth rate, GDP, and province-specific year trends are included in the estimation. Panel B replicates Panel A but uses county-specific year trends instead of province-specific year trends. Panel G uses the cohort shrinkage index as an alternative measure.  Robust standard errors, clustered at the county level, are reported in parentheses. *** $p<0.01$, ** $p<0.05$, * $p<0.1$.
 }
  \end{table}

Finally, we test the robustness of our famine intensity measure. In Panel G, we use an alternative famine measure: the Cohort Size Shrinkage Index (CSSI) constructed from the 1990 Census.\footnote{Following Huang et al. (2013), we use the 1990 Census (1\% sample) to construct a cohort-size shrinkage index, defined as the ratio of cohort size in a famine year to the average cohort size in non-famine years. The average cohort size for non-famine years is calculated over 1954-1957 and 1963-1965.} The CSSI provides an important cross-check because it is constructed from census-based cohort sizes rather than administrative mortality records, and is therefore not subject to the potential reporting biases in official death statistics discussed in Section 4. Although
the CSSI is subject to greater measurement error due to small county-level cohort sizes,\footnote{Although aggregating the CSSI over the three-year famine period could potentially address
the issue of small cohort sizes in a county-cohort cell, this approach is also likely to introduce
measurement error, since famine intensity varies substantially across famine years.} which
may attenuate the estimated effects, the results remain qualitatively consistent with our
baseline findings.

\subsection{Biological Scarring vs. Parental Behavioral Responses}
In addition to addressing gender-specific mortality selection, GS-SFE also accounts for time-invariant within-family gender disparities arising from parental gender preferences and cultural norms. This is particularly important given that son preference is well-documented phenomenon in China, implying differential treatment of daughters and sons within families. A central question in interpreting our findings is whether the estimated effects reflect biological scarring or postnatal behavioral responses by parents.\footnote{Because our sample includes only individuals born before the introduction of ultrasound technology in the 1980s-which enabled prenatal sex identification---concerns about gender-differentiated prenatal treatment are mitigated.} In particular, one may be concerned that parental gender preferences---and thus intra-household resource allocation---may vary even among siblings of the same gender, or may have temporarily shifted during the famine, thereby potentially bias our results.

We provide several pieces of evidence addressing this concern. First, parental treatment may differ among same-gender siblings. Elder siblings, especially the first son, might have received preferential treatment during the famine, potentially causing differences in outcome with other same gender siblings. To address this issue,  we explicitly control for birth order fixed effects and the number of older brothers and sisters in Table 6.  If the famine intensified son preference (e.g., prioritizing the eldest male), our estimates should change when controlling for intra-household hierarchy. The stability of our results across these specifications suggests that biological scarring, rather than shifts in postnatal parental preferences, is the dominant mechanism. 

Additionally, we test for compensatory investment behavior. Previous literature suggests that highly educated mothers are more likely to engage in compensatory investments to mitigate early-life shocks \citep{yi2015EJ,leightLiu2020EDCC}. In Panel D, we exclude mothers with a middle school education or higher. The estimates remain virtually unchanged. Given that this restriction removes a small fraction of the sample, it suggests that the results are driven by the general population of rural households rather than the specific behaviors of a higher-SES minority.

Second, we argue that even if parental treatment temporarily shifted during the famine, as long as same-gender siblings experienced the same shift and the impact of parental treatment is constant across ages, comparing a famine-born individual with their same-gender older sibling allows us to difference out the impact of such changes.\footnote{Note that comparisons with younger siblings cannot address this issue, because younger siblings did not experience the same shift in parental treatment.} Panel E restricts the sample to famine-cohort individuals and their older siblings by dropping individuals born after 1962. The estimation results remain similar to our main results, suggesting that concern over a shift in parental resource allocation is likely to be limited.

Third,  the lack of significant effects for postnatal exposure (ages 1-5) suggests that the damage is primarily driven by in utero deprivation rather than postnatal food allocation. As reported in Column 8 of Table 2, postnatal famine exposure (ages 1-5) has no statistically significant effect on educational outcomes. Moreover, if parents' son preference intensified during the famine relative to the non-famine period,  the estimated negative impact on males should be attenuated (biased toward zero). Yet we find large, negative, and statistically significant effects for males after correcting for mortality selection.

Finally, we examine fertility responses in Appendix Table \ref{tab:appendix_fertility}. If parents responded to the famine by altering their stopping rules---for example, deciding to stop having children after a famine-born son survived---this would introduce selection bias. We find no significant association between having a child during the famine and the total number or gender composition of subsequent children.

Taken together, these results suggest that while parental behaviors are undoubtedly complex, they do not appear to be the primary mechanism driving the observed educational deficits. The persistence of the effects across these robustness checks supports the interpretation that biological scarring and gender-specific mortality selection, rather than shifts in postnatal parental preferences, are the dominant forces at play.

\begin{table}
    \centering
    \caption{Biological Scarring vs. Parental Behavioral Responses.}
    \label{tab:rb2}
    \small
\begin{tabular}{lccc}
\toprule

\multicolumn{2}{c}{Years of education} & \multicolumn{2}{c}{Illiterate} \\
(1) & (2) & (3) & (4) \\
Prenatal EDR & Prenatal EDR$\times$Male & Prenatal EDR & Prenatal EDR$\times$Male \\
\midrule 

\multicolumn{4}{c}{\textbf{\textit{Panel A. Control for birth order FE}}} \\
$-5.848^{***}$ & 1.394 & $0.828^{***}$ & -0.267 \\
(1.915) & (2.573) & (0.247) & (0.304) \\

\multicolumn{4}{c}{\textbf{\textit{Panel B. Control for number of elder brothers}}} \\
$-5.799^{***}$ & 1.175 & $0.824^{***}$ & -0.259 \\
(1.895) & (2.495) & (0.253) & (0.305) \\

\multicolumn{4}{c}{\textbf{\textit{Panel C. Control for number of elder sisters}}} \\
$-5.718^{***}$ & 1.067 & $0.811^{***}$ & -0.239 \\
(1.843) & (2.469) & (0.254) & (0.311) \\

\multicolumn{4}{c}{\textbf{\textit{Panel D. Drop mother with higher education}}} \\
$-5.827^{***}$ & 0.677 & $0.821^{***}$ & -0.241 \\
(1.864) & (2.540) & (0.251) & (0.305) \\
\multicolumn{4}{c}{\textbf{\textit{Panel E. Drop individuals born after 1962}}} \\
    -4.358* & -2.734 & 0.699** & 0.050 \\
    (2.349)   & (3.458) & (0.315)   & (0.490) \\
\bottomrule
\end{tabular}

\caption*{ 
\footnotesize Notes: Gender-specific birth year fixed effects, gender-specific sibling fixed effects, and EDR at ages 1-5 are controlled for in all columns. Each panel reports a single regression, separately estimated using Years of Education and Illiterate as dependent variables, respectively. The reported coefficients reflect the effect of prenatal exposure to famine on both genders (the coefficient on Prenatal EDR), and the differential effect for males relative to females (the coefficient on $Prenatal\ EDR \times Male$), as indicated in the top panel. The sample size for Panels A to C is 4,055 while the sample size for Panel D and E is 3,927, and 1043 respectively. In Panel D, a mother with higher education is defined as having completed at least middle school.  Robust standard errors, clustered at the county level, are reported in parentheses. *** $p<0.01$, ** $p<0.05$, * $p<0.1$.
 }
  \end{table}

\section{Conclusion}

This study examines the enduring consequences of prenatal exposure to the Great Chinese Famine on educational outcomes. By addressing \textit{gender-specific selection bias}, our research helps resolve a persistent puzzle in the fetal origins literature: why males, despite being biologically more vulnerable to early-life shocks, often appear less affected in long-term observational studies.

We employed an approach of comparing famine-exposed individuals with their unexposed \textit{same-gender} siblings. This framework effectively mitigates the selection bias stemming from the ``fragile male'' phenomenon---where higher mortality thresholds for males censor the most affected individuals from the sample. Our estimates show that the long-term scarring effects are remarkably similar across genders: both males and females experienced an increase in illiteracy rates by 4 percentage points and a decrease in years of education by 0.3 years. Crucially, these estimated negative effects on males are larger than those derived from standard models, indicating that standard approaches underestimate fetal shocks by overlooking gender-specific mortality.

Furthermore, our analysis of mechanisms indicates that these deficits are driven primarily by biological scarring in utero rather than postnatal behavioral responses. We find no significant evidence that the results are confounded by changes in fertility stopping rules, sibling composition, or compensatory parental investments.

Methodologically, this paper demonstrates the utility of Gender-Specific Sibling Fixed Effects (GS-SFE) in contexts characterized by high fertility, where the trade-off between bias reduction and estimation efficiency is manageable. Our findings emphasize the critical importance of maternal nutrition during pregnancy and highlight the potential for intergenerational transmission of disadvantage through reduced human capital.

 \pagebreak{}

\noindent \singlespacing

\pagebreak
\doublespacing
\pagestyle{empty}

\renewcommand{\theequation}{A.\arabic{equation}}
\setcounter{equation}{0}
\appendix

\section{Sample Selection Bias}
\label{app:A}

When $\gamma_{g}<0$ holds and $E[u_{ig}|v_{ig}]$ is an increasing
function of $v_{ig}$, then $E[u_{ig}|v_{ig}\geq-\gamma_{0}-\gamma_{g}EDR_{i}]$
is increasing in $EDR_{i}$. This means that, if we estimate (\ref{eq:main})
by the OLS using the selected sample of $S_{ig}=1$, the OLS estimator
for $\beta_{g}$ will be upwardly biased, where the asymptotic bias
is given by

\begin{equation}\label{eq:bias}
\hat{\beta}_{g}-\beta_{g}\overset{p}{{\rightarrow}}\frac{{Cov(EDR_{i,}E[u_{i}|v_{i}\geq-\gamma_{0}-\gamma_{g}EDR_{i}])}}{Var(EDR_{i})}.
\end{equation}
The magnitude of this sample selection bias is expected to be increasing
in the value of $|\gamma_{g}|$.  When \(u_{i}\) and \(v_{i}\) are jointly normally distributed, we have a Heckman's sample selection model with
$E[u_{i} \mid v_{i} \geq -\gamma_{0} - \gamma_{g}EDR_{i}] = \sigma_{uv}\lambda(\gamma_{0} + \gamma_{g}EDR_{i})$,
where \(\lambda(z) := \frac{\phi(z)}{\Phi(z)}\) is the inverse Mills ratio with \(\lambda'(z) > 0\). We expect \(\sigma_{uv} = \text{Cov}(u,v) > 0\) if unobserved individual traits both  enhance survival probability and promote better educational achievement. Then, when \(\gamma_{g} < 0\), \begin{equation}\label{selection}
\text{Cov}(EDR_{i}, E[u_{i} \mid v_{i} \geq -\gamma_{0} - \gamma_{g}EDR_{i}]) = \sigma_{uv}\text{Cov}(EDR_{i},\lambda(\gamma_{0} + \gamma_{g}EDR_{i}))< 0,
\end{equation}
and
the estimated effect of prenatal famine exposure on educational attainment, $\hat\beta_g$, will be negatively biased.

Furthermore, if male and female share
the same parameter values for $\sigma_{uv}$ and $\gamma_{0}$, but
the effect of EDR on death in early childhood is larger for male than
female so that $\gamma_{\text{male}}<\gamma_{\text{female}}<0$, then
this sample selection bias is larger for male than for female because
the covariance between $EDR_{i}$ and $\sigma_{uv}\lambda(\gamma_{0}+\gamma_{g}EDR_{i})$
is larger for male than for female.

\section{Gender-specific Sibling Fixed Effects and Derivation of Equation (\ref{eq:ATT-S})}
  
\label{app:B}
\renewcommand{\theequation}{B.\arabic{equation}}
\setcounter{equation}{0}
Consider the same-gender sibling of individual $i$, denoted by $i'$, who was born outside the famine period such that $EDR_{i'}=0$. For brevity, we assume that $(\epsilon_i,\eta_i)$ is independent of $(\epsilon_{i'},\eta_{i'})$ conditional on $\xi_{j(i)g}$ and that  $\epsilon_{i'}$ is mean independent of $\eta_{i'}$ given $\xi_{j(i)g}$, i.e., sibling $i'$ who was born outside of the famine period is not subject to selection issue. Then, following equations (\ref{eq:main}) and (\ref{eq:decomposition}), taking the conditional expectation of the difference between $Y_{ig}$ and $Y_{i'g}$ given $S_{ig}=S_{i'g}=1$ and $EDR_{i}$  under the assumption that  (i) $(\epsilon_{i},\eta_{i})$ is conditionally independent of $(\epsilon_{i'},\eta_{i'})$ given $\xi_{jg}$ and that (ii) $E[ \epsilon_{i'}| \eta_{i'},\xi_{jg}]=0$, for $j=j(i)=j(i')$, we obtain  
\begin{align}\label{eq:main-12}
E[Y_{ig}-Y_{i'g} | EDR_{i}, S_{ig} =S_{i'g}=1] &= \beta_{g} EDR_{i} +E[ \epsilon_{i}| \eta_{i} \geq -\gamma_{0} - \gamma_{g} EDR_{i} - \xi_{{jg}}].
\end{align}  
because
\begin{align*}
&E[ \epsilon_{i}-\epsilon_{i'} | \eta_{i} \geq -\gamma_{0} - \gamma_{g} EDR_{i} - \xi_{{jg}},\eta_{i'} \geq -\gamma_{0} - \xi_{{jg}}]\\
=&E[ \epsilon_{i}  | \eta_{i} \geq -\gamma_{0} - \gamma_{g} EDR_{i} - \xi_{{jg}}]
-E[  \epsilon_{i'} | \eta_{i'} \geq -\gamma_{0} - \xi_{{jg}}]\\
=&E[ \epsilon_{i}  | \eta_{i} \geq -\gamma_{0} - \gamma_{g} EDR_{i} - \xi_{{jg}}],
\end{align*}
where the first equality assumes $EDR_{i'}=0$, the second equality follows from the independence between $(\epsilon_{i},\eta_{i})$ and $(\epsilon_{i'},\eta_{i'})$, and  the last equality follows from the Law of Iterated Expectations with $E[ \epsilon_{i'}| \eta_{i'},\xi_{jg}]=0$. Given the assumption of no selection issue arising from sibling $i'$, for brevity, we subsume the conditioning variable $S_{i'g}=1$.

Comparing equations (\ref{eq:condieq}) and (\ref{eq:main-12}), and considering the asymptotic selection bias of the form (\ref{eq:bias}), we find that controlling for \textit{gender-specific} sibling fixed effects reduces selection bias when gender-specific unobserved family characteristics are the predominant source of such bias ---that is, when $\text{Cov}[ \alpha_{jg} + \epsilon_{i}, \xi_{jg} + \eta_{i}]$ is positive and substantially larger than $\text{Cov}[ \epsilon_{i}, \eta_{i}]$. If $\epsilon_i$ is mean independent of $\eta_i$ and $\xi_{{jg}}$, then the gender-specific family fixed effects approach consistently estimates $\beta_g$ even when $\alpha_{jg}$ and $\xi_{ig}$ are correlated.

When we use the binary treatment variable $D_i$ in place of $EDR_i$, for sibling pairs $(i,i')$ of the same gender $g$, we have:
   \begin{equation*}
   Y_{ig} - Y_{i'g} = \beta_{ig}D_i + (\epsilon_i - \epsilon_{i'}),
   \end{equation*}
   where  $D_{i'} =0$. Regressing $Y_{ig} - Y_{i'g}$ on $D_i$ using the survivor's sample, the probability limit of  the gender-specific family fixed effects estimator $ \hat{\beta}_{g, FE} $ can be written as:
  \begin{align*}
 \hat{\beta}_{g, FE}& \overset{p}{\rightarrow} \frac{E[(D_{i}-E[D_i|S_{ig}=1])(Y_{ig} - Y_{i'g})| S_{ig}=1] }{ Var(D_i|S_{ig}=1)}\\
 &= \frac{E[(D_{i}-E[D_i|S_{ig}=1])( \beta_{ig}D_i + (\epsilon_i - \epsilon_{i'}))|D_i=1, S_{ig}=1] }{ Var(D_i|S_{ig}=1)}\\
 &= \frac{E[(D_{i}-E[D_i|S_{ig}=1])D_i  \beta_{ig} | S_{ig}=1] }{ Var(D_i|S_{ig}=1)}\\
 &= E[\beta_{ig} | D_i=1,S_{ig}=1],
 \end{align*}
 where the third equality assumes that $\epsilon_i$ and $\epsilon_{i'}$ are conditionally mean independent of $\eta_i$ and $\eta_{i'}$ given $\xi_{{jg}}$ while the last line follows 
because the numerator is written as
\begin{align*}
&E[(D_{i}-E[D_i|S_{ig}=1])D_i  \beta_{ig} | S_{ig}=1]\\
=& E[(1-E[D_i|S_{ig}=1]) \beta_{ig} | D_i=1,S_{ig}=1]\Pr(D_i=1|S_{ig}=1)\\
&\quad + E[(0-E[D_i|S_{ig}=1]) \times 0\times \beta_{ig} | D_i=0,S_{ig}=1]\Pr(D_i=0|S_{ig}=1)\\
&= E[\beta_{ig} | D_i=1,S_{ig}=1] \Pr(D_i=1|S_{ig}=1)(1-\Pr(D_i=1|S_{ig}=1)).
\end{align*}
 while $Var(D_i|S_{ig}=1)=\Pr(D_i=1|S_{ig}=1)(1-\Pr(D_i=1|S_{ig}=1))$. Therefore, (\ref{eq:ATT-S}) holds.

\section{Bias-Efficiency Trade-off and Sample Representativeness}
\label{app:C}

While the GS-SFE model addresses the critical issue of gender-differentiated selection bias, it introduces two practical trade-offs: estimation efficiency and sample representativeness.

First, regarding efficiency, standard sibling fixed effects models utilize variation from all multi-child families. In contrast, the GS-SFE approach relies exclusively on within-family variation among siblings of the same gender. Consequently, ``singleton-gender'' siblings---for example, a male with only female siblings---do not contribute to the identification of the male-specific treatment effect ($\beta_{male}$). This restriction reduces the effective degrees of freedom, potentially inflating standard errors. As shown in Table 1, even after restricting identification to same-gender comparisons, we retain a substantial number of effective sibling pairs (over 2,000). To make this trade-off transparent, we report the effective number of observations in our regression results.

Second, a related concern is whether the ``effective sample'' contributing to the GS-SFE identification is representative of the general population. Identification strictly comes from families with at least two children of the same gender (e.g., two boys or two girls). This restriction mechanically selects for larger families. In many empirical settings, larger families are associated with specific characteristics, eg, parental income or education.

However, in the context of rural China during this period, this concern is mitigated by the universally high fertility rates driven by post-war population expansion policies. As noted, the median family size in our sample is four. Consequently, families with same-gender sibling pairs are not statistical outliers but rather the norm. To verify this,  we compared the parental education levels (a key proxy for family SES) between families that contribute to our identification (those with same-gender sibling pairs) and those that do not. As detailed in Table \ref{tab:balance_check}), we find no statistically significant differences in either father's or mother's years of schooling between these two groups. This provides suggestive evidence that families in our GS-SFE is likely to be representative. 
Furthermore, to ensure our results are not driven by the specific demographic composition of the effective sample, we performed a robustness check using Inverse Probability Weighting (IPW). We re-weighted the effective sample to match the probability of inclusion based on birth year and county characteristics. As reported in Table \ref{tab:reweighting}, the estimated treatment effects remain quantitatively similar to our baseline results, suggesting that the concern over the representativeness of the sample is likely to be limited.

\setcounter{table}{0}
\renewcommand{\thetable}{C\arabic{table}}

\begin{table}[htbp]
\centering
\caption{Balance Check: Parental Education by Family Structure}
\label{tab:balance_check}
\begin{tabular}{lccc}
\hline \hline
 & (1) & (2) & (3) \\
 & Families with & Families without & Difference \\
 & Same-Gender Pairs & Same-Gender Pairs & (t-stat) \\
\hline
\textbf{Father's Years of Education} & 3.44 & 3.70 & -0.26 \\
 & (0.13) & (0.31) & (-0.772) \\
 & & & \\
\textbf{Mother's Years of Education} & 1.28 & 1.47 & -0.19 \\
 & (0.09) & (0.22) & (-0.866) \\
\hline \hline
\end{tabular}
\begin{minipage}{0.9\textwidth} 
\footnotesize \vspace{0.1cm}
\textit{Notes:} Sample restricted to individuals born before 1972. Families ``without same-gender pairs'' primarily consist of two-child families with one boy and one girl. Standard errors of means in parentheses.
\end{minipage}
\end{table}
\begin{table}[htbp]
\centering
\caption{Robustness Check: Inverse Probability Weighting (IPW)}
\label{tab:reweighting}
\begin{tabular}{lcccc}
\hline \hline
 & \multicolumn{2}{c}{Years of Education} & \multicolumn{2}{c}{Illiterate} \\
 \cline{2-3} \cline{4-5}
 & (1) & (2) & (3) & (4) \\
 & Baseline & Re-weighted (IPW) & Baseline & Re-weighted (IPW) \\
\hline
Prenatal EDR & -5.795*** & -6.420*** & 0.829*** & 0.820*** \\
 & (1.612) & (1.664) & (0.283) & (0.285) \\
 & & & & \\
Prenatal EDR $\times$ Male & 1.444 & 0.611 & -0.241 & -0.146 \\
 & (2.342) & (2.380) & (0.326) & (0.331) \\
 & & & & \\
\hline
Gender-specific Sibling FE & Yes & Yes & Yes & Yes \\
Gender-specific Birth Year FE & Yes & Yes & Yes & Yes \\
Observations & 4,155 & 4,125 & 4,155 & 4,125 \\
\hline \hline
\end{tabular}
\begin{minipage}{0.9\textwidth} 
\footnotesize \vspace{0.1cm}
\textit{Notes:} Columns (1) and (3) report the baseline estimates from Table 2 and 3 (Column 7). Columns (2) and (4) report estimates where sample weights are applied. We calculated the probability that an individual from the full sample is included in the GS-SFE model conditional on their birth year and county. We then applied the inverse of this predicted probability as a weight in the regression.  Standard errors, clustered at the county level, are reported in parentheses. *** p$<$0.01, ** p$<$0.05, * p$<$0.1.
\end{minipage}
\end{table}


\section{Gender-Specific Famine Intensity Measures and Attenuation Bias}
\label{app:D}
\renewcommand{\theequation}{D.\arabic{equation}}
\setcounter{equation}{0}

In our main analysis, we utilize a gender-neutral Excess Death Rate (EDR) to measure famine intensity. While one might argue that gender-specific EDRs (if available) could better capture the specific exposure experienced by males and females, we argue that using gender-specific mortality rates as a proxy for famine intensity introduces a mechanical bias that attenuates the estimated effects for males.

Conceptually, the goal of the identification strategy is to estimate the impact of an environmental shock (e.g., severe caloric restriction) on fetal development. The ``fragile male'' hypothesis posits that for the exact same level of nutritional deprivation, males experience higher mortality rates than females \citep{Kraemer2000}. Consequently, if gender-specific EDRs were used, the measure for males would be systematically higher than for females for the same underlying famine intensity. Using a male-specific EDR would inflate the denominator of the estimated coefficient, mechanically reducing the magnitude of the effect relative to females, even if the biological scarring response to the nutritional shock is identical.

A formal argument is given as follows. Let $F_i^*$ denote the latent famine intensity (e.g., the degree of caloric deprivation) experienced by the mother of individual $i$. The structural relationship of interest is:
\begin{equation}
    Y_{ig} = \alpha + \beta_g F_i^* + \epsilon_{ig},
\end{equation}
where $Y_{ig}$ is the outcome (e.g., education) for gender $g \in \{m, f\}$, and $\beta_g$ captures the true biological scarring effect of the nutritional shock on gender $g$.

The gender-specific Excess Death Rate ($EDR_{ig}$) is a biological outcome of this shock, not the shock itself. We model this relationship as:
\begin{equation}
    EDR_{ig} = \lambda_g F_i^*
\end{equation}
where $\lambda_g$ represents gender-specific mortality sensitivity. The ``fragile male'' hypothesis implies that males have higher mortality sensitivity to the same shock, such that:
\begin{equation}
    \lambda_m > \lambda_f > 0.
\end{equation}
If we were to use gender-specific $EDR_{ig}$ as the regressor in place of the unobserved $F_i^*$, we would estimate the following equation:
\begin{equation}
    Y_{ig} = \alpha + \tilde{\beta}_g EDR_{ig} + u_{ig}.
\end{equation}
Substituting $F_i^* = EDR_{ig} / \lambda_g$ into the structural equation, the probability limit of the OLS estimator $\tilde{\beta}_g$ is given by:
\begin{equation}
    \tilde{\beta}_g = \frac{\beta_g}{\lambda_g}.
\end{equation}
This derivation highlights the bias from using gender-specific EDR. Even if the true scarring effects are identical across genders (i.e., $\beta_m = \beta_f = \beta < 0$), the estimated coefficients will differ mechanically due to the difference in $\lambda_g$. Specifically, since $\lambda_m > \lambda_f$, it follows that:
\begin{equation}
    |\tilde{\beta}_m| = \left| \frac{\beta}{\lambda_m} \right| < \left| \frac{\beta}{\lambda_f} \right| = |\tilde{\beta}_f|.
\end{equation}
Thus, using gender-specific EDRs would mechanically attenuate the estimated effect for males relative to females ($|\tilde{\beta}_m| < |\tilde{\beta}_f|$), leading to spurious evidence of gender differences. By contrast, using a gender-neutral EDR proxies for $F_i^*$ directly without scaling by gender-specific mortality sensitivity, thereby minimizing this measurement bias.

\section{Additional Tables and Figures}

\begin{figure*}
\center
\includegraphics[scale=1]{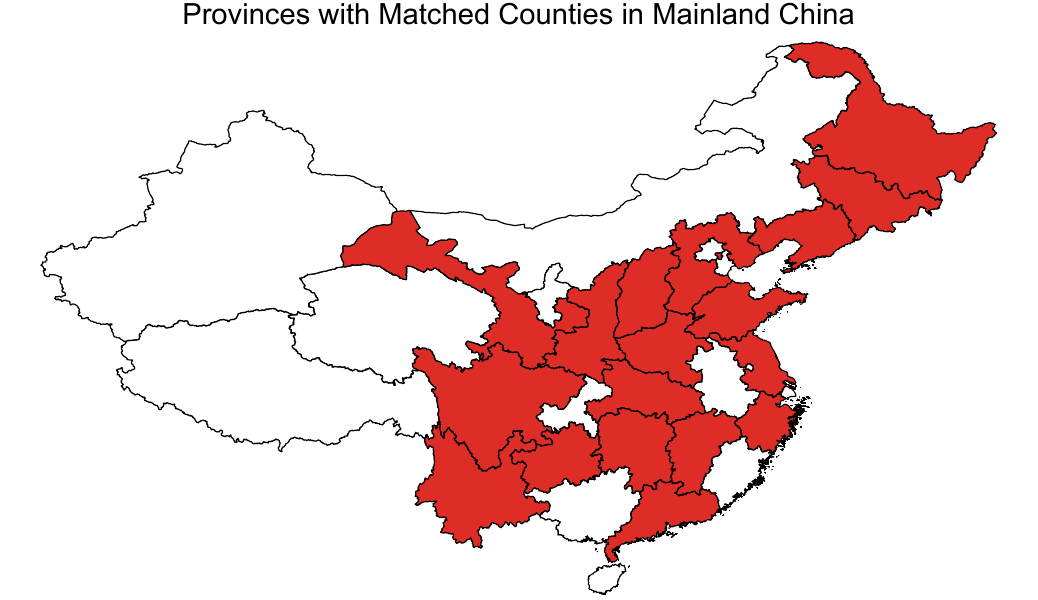} 
\end{figure*}
 
\clearpage
\setcounter{table}{0} 
\renewcommand{\thetable}{E\arabic{table}} 

\begin{table}
\centering
\caption{The Long-Term Effects of Prenatal Exposure to the Chinese Famine, Using a Binary Treatment Variable}
\label{tab:appendix_binary}
\begin{tabular}{lcccc}
\toprule
 & (1) & (2) & (3) & (4) \\
\midrule
\multicolumn{5}{l}{\textbf{Panel A. Years of Education}} \\

Prenatal EDR Dummy 
    & -1.425** &        & -1.322** & -1.227* \\
    & (0.545)  &        & (0.612)  & (0.643) \\

Prenatal EDR Dummy $\times$ Male 
    &          & -1.548** & -0.226 & -0.310 \\
    &          & (0.713)  & (0.745) & (0.758) \\

Prenatal EDR Dummy $\times$ Female
    &          & -1.322** &         &  \\
    &          & (0.612)  &         &  \\

\multicolumn{5}{l}{\textbf{Panel B. Illiterate}} \\

Prenatal EDR Dummy 
    & 0.178*** &          & 0.182*** & 0.180*** \\
    & (0.054)  &          & (0.062)  & (0.067) \\

Prenatal EDR Dummy $\times$ Male 
    &          & 0.172** & -0.010 & -0.007 \\
    &          & (0.071) & (0.077) & (0.077) \\

Prenatal EDR Dummy $\times$ Female
    &          & 0.182*** &         & \\
    &          & (0.062)  &         & \\
 \midrule
Observations & 4,155 & 4,155 & 4,155 & 4,055 \\
\bottomrule
\end{tabular}
\caption*{Notes: Model specifications in Panels A and B are the same as Columns 5-8 of Table 2 and Table 3. The reported observations refer to the effective sample size contributing to identification (i.e., excluding singleton observations dropped by the fixed effects). Standard errors, clustered at the county level, are reported in parentheses. *** $p<0.01$, ** $p<0.05$, * $p<0.1$.}
\end{table}

\begin{table}
\centering
\caption{ Drop Individuals Born 1953-1955}
\label{tab:appendix_drop1953}
\begin{tabular}{lcc}
\toprule
 & (1) & (2) \\
 & Years of education & Illiterate \\
\midrule
Prenatal EDR &  $-6.131^{***}$ & $0.830^{***}$ \\
               & $(2.117)$      & $(0.283)$      \\

Prenatal EDR*Male & $1.875$        & $-0.236$       \\
               & $(2.603)$      & $(0.302)$      \\

\midrule
Observations   & $3,469$        & $3,469$        \\
R-squared      & $0.719$        & $0.692$        \\ \hline
\bottomrule
\end{tabular}
\caption*{
\small
 Note: Model specification is the same as the last column of Table 2 and 3. 
Standard errors, clustered at the county level, are reported in parentheses. 			
*** $p<0.01$, ** $p<0.05$, * $p<0.1$.}

\end{table}

\begin{table}[htbp]
    \centering
    \caption{Coefficients on Triple-Interaction Terms: Prenatal EDR in 1960 $\times$ Male $\times$ Cohort Dummy}
    \label{tab:triple_interaction}
    \footnotesize
\begin{tabular}{lcccc}
\toprule
 & \multicolumn{2}{c}{Years of education} & \multicolumn{2}{c}{Illiterate} \\
\cmidrule(lr){2-3} \cmidrule(lr){4-5}
 & (1) & (2) & (3) & (4) \\
\midrule

\multicolumn{5}{l}{\textit{Pre-famine cohorts: Prenatal EDR in 1960 $\times$ Male $\times$ Cohort}} \\
1949--1952 & $3.780$ & $3.913$ & $-0.068$ & $-0.083$ \\
 & $(3.055)$ & $(3.104)$ & $(0.352)$ & $(0.352)$ \\
1953--1955 & $-0.641$ & $-0.541$ & $0.019$ & $0.008$ \\
 & $(1.820)$ & $(1.823)$ & $(0.152)$ & $(0.151)$ \\
1956--1958 & $3.020$ & $3.208$ & $-0.357$ & $-0.377$ \\
 & $(2.373)$ & $(2.392)$ & $(0.263)$ & $(0.261)$ \\
\multicolumn{5}{l}{\textit{Post-famine cohorts:  Prenatal EDR in 1960 $\times$ Male $\times$ Cohort}} \\
1966--1968 & $1.899$ & $2.037$ & $0.013$ & $-0.002$ \\
 & $(1.652)$ & $(1.668)$ & $(0.123)$ & $(0.126)$ \\
1969--1971 & $-0.137$ & $-0.010$ & $-0.029$ & $-0.043$ \\
 & $(1.359)$ & $(1.381)$ & $(0.115)$ & $(0.119)$ \\
1972--1974 & $1.272$ & $1.388$ & $-0.090$ & $-0.103$ \\
 & $(0.979)$ & $(1.000)$ & $(0.116)$ & $(0.121)$ \\
1975--1979 & $0.134$ & $0.238$ & $-0.005$ & $-0.016$ \\
 & $(1.880)$ & $(1.890)$ & $(0.175)$ & $(0.171)$ \\

\midrule
Observations & 3,711 & 3,711 & 3,711 & 3,711 \\
\bottomrule
\end{tabular}
\caption*{\footnotesize
Note: This table reports the coefficients on the triple-interaction terms (Prenatal EDR in 1960 $\times$ Male $\times$ Cohort Dummy) estimated in Table \ref{tab:trend}.  Robust standard errors, clustered at the county level, are reported in parentheses. *** $p<0.01$, ** $p<0.05$, * $p<0.1$.
}
\end{table}

\begin{table}[htbp]
\centering
\caption{Test whether having a child born during the famine could reduce total family size}
\label{tab:appendix_fertility}
\begin{tabular}{lcccccc}
\toprule
& \multicolumn{2}{c}{\shortstack{Having at least one\\child born after  famine}} 
& \multicolumn{2}{c}{\shortstack{Number of children\\born after famine }} 
& \multicolumn{2}{c}{\shortstack{Gender ratio among\\children born after famine}} \\
& (1) & (2) & (3) & (4) & (5) & (6) \\
\midrule
Had child born & 0.002 & & 0.041 & & -0.018 & \\
\ in famine& (0.029) & & (0.114) & & (0.123) & \\
Male child born & & -0.006 & & -0.109 & & -0.096 \\
\ in famine& & (0.035) & & (0.136) & & (0.164) \\
Female child born & & -0.006 & & 0.052 & & 0.038 \\
\ in famine& & (0.035) & & (0.150) & & (0.123) \\
\\
\midrule
Observations & 2,876 & 2,876 & 2,878 & 2,878 & 1,732 & 1,732 \\
R-squared & 0.288 & 0.288 & 0.341 & 0.342 & 0.103 & 0.106 \\
\bottomrule
\end{tabular}

\caption*{\small 
Note: Sample is restricted to families that have at least one child born by the end of the famine. All columns control for the number of male and female children born before the end of the famine, birth year fixed effects of the first child, gender of the first child, and county fixed effects. Standard errors, clustered at the county level, are reported in parentheses. *** $p<0.01$, ** $p<0.05$, * $p<0.1$.
}
\end{table}

\end{document}